\documentclass[
	a4paper, 
	10pt, 
]{LTJournalArticle}

\usepackage{calc}
\usepackage{amsthm}
\usepackage{dsfont}
\usepackage{xcolor}
\usepackage{pslatex}
\usepackage{apalike}
\usepackage{amssymb}
\usepackage{amstext}
\usepackage{amsmath}
\usepackage{multicol}
\usepackage{stmaryrd}
\usepackage{hyperref}
\usepackage{dirtytalk}
\usepackage{subcaption}
\usepackage[bottom]{footmisc}
\usepackage[linesnumbered,ruled]{algorithm2e}

\newtheorem{proposition}{Proposition}

\newtheorem{corollary}{Corollary}
\newtheorem{criterion}{Criterion}

\newtheorem{definition}{Definition}

\hypersetup{
    pdftoolbar=true,        
    pdfmenubar=true,        
    pdffitwindow=false,     
    pdfstartview={FitH},    
    pdftitle={Euclidean Distance to Convex Polyhedra and Application to Class Representation in Spectral Images},    
    pdfauthor={Antoine Bottenmuller} {Florent Magaud} {Arnaud Demortière} {Etienne Decencière} {Petr Dokladal},     
    pdfsubject={Mathematics and Machine Learning for Spectral Unmixing},   
    pdfcreator={Antoine Bottenmuller},   
    pdfproducer={Antoine Bottenmuller}, 
    pdfkeywords={Spectral Image} {Linear Unmixing} {Abundance Map} {Density Function} {Linear Classifier} {Convex Polyhedron}, 
    pdfnewwindow=true,      
    colorlinks=true,       
    linkcolor=[rgb]{0.9,0,0},    
    citecolor=[rgb]{0,0.7,0},    
    filecolor=[rgb]{1,0,1},    
    menucolor=[rgb]{0,1,1},    
    urlcolor=[rgb]{0,0,0.7},    
}
\definecolor{orcidColor}{rgb}{0.0, 0.0, 0.0}
\definecolor{emailColor}{rgb}{0.0, 0.0, 0.0}

\title{Euclidean Distance to Convex Polyhedra and\\Application to Class Representation in Spectral Images} 

\author{%
	\href{https://orcid.org/0009-0007-3943-3419}{\textcolor{orcidColor}{Antoine Bottenmuller}}\textsuperscript{1}, \href{https://orcid.org/0009-0009-1817-4408}{\textcolor{orcidColor}{Florent Magaud}}\textsuperscript{2}, \href{https://orcid.org/0000-0002-4706-4592}{\textcolor{orcidColor}{Arnaud Demortière}}\textsuperscript{2},\\\href{https://orcid.org/0000-0002-1349-8042}{\textcolor{orcidColor}{Etienne Decencière}}\textsuperscript{1} and \href{https://orcid.org/0000-0002-6502-7461}{\textcolor{orcidColor}{Petr Dokladal}}\textsuperscript{1}
}

\date{
    \footnotesize\textsuperscript{\textbf{1}}Center for Mathematical Morphology (CMM), Mines Paris, PSL University, Fontainebleau, France \\
    \textsuperscript{\textbf{2}}Laboratoire de Réactivité et Chimie des Solides (LRCS), CNRS UMR 7314, Université de Picardie Jules Verne, Amiens, France \\
    \vspace{0.80\baselineskip}
    \{\textit{\href{mailto:antoine.bottenmuller@minesparis.psl.eu}{\textcolor{emailColor}{antoine.bottenmuller}}, \href{mailto:etienne.decenciere@minesparis.psl.eu}{\textcolor{emailColor}{etienne.decenciere}}, \href{mailto:petr.dokladal@minesparis.psl.eu}{\textcolor{emailColor}{petr.dokladal}}}\}@\textit{minesparis.psl.eu} \\
    \{\textit{\href{mailto:florent.magaud@u-picardie.fr}{\textcolor{emailColor}{florent.magaud}}, \href{mailto:arnaud.demortiere@u-picardie.fr}{\textcolor{emailColor}{arnaud.demortiere}}}\}@\textit{u-picardie.fr}
}

\renewcommand{\maketitlehookd}{%
	\begin{abstract}
		\noindent With the aim of estimating the abundance map from observations only, linear unmixing approaches are not always suitable to spectral images, especially when the number of bands is too small or when the spectra of the observed data are too correlated. To address this issue in the general case, we present a novel approach which provides an adapted spatial density function based on any arbitrary linear classifier. A robust mathematical formulation for computing the Euclidean distance to polyhedral sets is presented, along with an efficient algorithm that provides the exact minimum-norm point in a polyhedron. An empirical evaluation on the widely-used Samson hyperspectral dataset demonstrates that the proposed method surpasses state-of-the-art approaches in reconstructing abundance maps. Furthermore, its application to spectral images of a Lithium-ion battery, incompatible with linear unmixing models, validates the method's generality and effectiveness.
	\end{abstract}
}

\begin{document}

\maketitle 

\noindent\textbf{\textit{Keywords }} 
Spectral Image $\cdot$ Linear Unmixing $\cdot$ Abundance Map $\cdot$ Density Function $\cdot$ Linear Classifier $\cdot$ Convex Polyhedron.
\vspace{-0.0\baselineskip}



\section{Introduction}
\label{sec:introduction}

\subsection{Context}

\noindent Spectral images have become a common type of data widely used in a large set of scientific domains and for various applications, such as agriculture for vegetation identification, materials science for defect detection, chemistry for compound quantification, or satellite imaging for geosciences or for a military usage. 

The general term of \say{spectral imaging} covers all imaging techniques where two or more spectral bands are used to capture the data: RGB (three bands), multispectral (three to tens of bands), hyperspectral (hundreds to thousands of continuous spectral bands) and multiband (spaced spectral bands) imaging. In such images, whether it is the absorption or the reflectance of the observed matter that is measured, to each pixel is associated one spectrum - or one vector of $n$ spectral band values -, which can be represented as one unique element in a $n$-dimensional Euclidean space, usually $\mathbb{R}^n$, with $n$ the number of spectral bands. 

In usual cases, pixels' spectra are considered as linear mixtures of pure class spectra, called endmembers. For example, in satellite imaging, a low spatial resolution produces mixtures of geographical areas ; or in industrial chemistry, the observed substances are mixtures of pure chemical compounds. The following matrix equation allows describing this modelling: 

\vspace{-1.5mm}
\begin{equation}
    Y = MA
    \label{eq:YMA}
\end{equation}
\vspace{-0.5mm}

\noindent where $Y$ is the matrix of the observed data of size $n \times k$ (for $n$ spectral bands and $k$ pixels), $M$ is the matrix of the endmembers of size $n \times m$ (for $m$ endmembers) where each column represents an endmember's spectrum, and $A$ the matrix of the abundances of size $m \times k$ representing the proportions of each endmember in the spectral composition of the pixels \cite{tao2021endmember}. Usually, a little Gaussian noise matrix $\epsilon$ is added to the equation (Eq.\ref{eq:YMA}). The main challenge is then to find back, from the observations $Y$ only, both the matrices $M$ and $A$: this process is called \textit{linear unmixing}.

Numerous recent methods have been developed over this linear-mixture modelling: geometrical approaches \cite{winter1999n}, variational inverse problems \cite{eches2011variational}, bayesian methods \cite{figliuzzi2016bayesian}, or deep learning approaches \cite{chen2023improved}. Most of them determining first the matrix $M$ of the endmembers, then reduce and invert $M$ to build back the abundance matrix $A$ as $A = M^{-1}Y$. 
Although the assertion of linear mixture of endmembers is faithful to physical reality, trying to recover the abundances $A$ from estimated endmembers $M$ can be sometimes either not sufficient, not pertinent or even impossible to achieve, especially when:

\begin{itemize}

    \item the number of spectral bands $n$ is lower than the number $m$ of considered classes (or endmembers), making $M$ not reducible and thus not invertible ;
    
    \item the captured spectra are too correlated to each other, which tends to make the determined endmembers linearly dependant, and therefore making $M$ not invertible ;
    
    \item or the observed mixed spectra are too far or isolated from the true endmembers (the mixture is too strong), and thus looking for the endmembers becomes hard or even not pertinent. 
    
\end{itemize}

Usual linear unmixing approaches may particularly be not appropriate for data captured under a few bands only (RGB, multispectral or multiband images), or to the cases where even the true endmembers are too correlated or linearly dependant (usually due to a lack of spectral bands captured), for the second reason above. An example of this last situation is used in the applications (Section \ref{sec:application}). In such cases, one more general way of estimating $A$ would be the use of image analysis or data segmentation approaches, allowing both classifying the data and determining a probability map that can be interpreted as the abundance map, by using an adapted density function, like for deep learning models or clustering algorithms. And this, without having to find the endmembers $M$.

In this article, we consider these last approaches, to be able to classify the data in the general case (whether the linear-mixture modelling is suitable or not), and we present a new and simple method which allows building an appropriate density map associated with the classes given by any arbitrary linear (or \textit{polyhedral}) classifier over spectral images.


\subsection{Objectives}

\noindent The main objective is then, given any spectral image, to build back - or give a good approximation of - the abundance map or the probability map associated with the observations, using a data segmentation approach for the general case compatibility and for a greater control over the classification. To achieve this, two successive processes must be predetermined:

\begin{enumerate}
    
    \item an arbitrarily-chosen classifier, which allows segmenting the Euclidean spectral space into distinct and complementary areas, each of them representing one of the computed classes ;
    
    \item the spatial density function, which allows, from the classification made over the data, computing a continuous spatial distribution (abundance or probability) of the classes in space.
    
\end{enumerate}

Note that the terms \say{abundance} and \say{probability} have a different conceptual interpretation: on the one hand, the abundance map represents the proportion of presence of each class in the observed pixels (considered as class mixtures), where, on the other hand, the probability map represents the associated probability of the observed pixels to belong to each of the classes. We use the word \say{density} to gather both terms.

Deep learning approaches allow getting density functions by extracting the last layer of the networks after the softmax function, or by taking values in their latent space. But they are often complex, over-parameterized, need prior information or a minimum of training, and the majority of the state-of-the-art architectures seem to produce poorer results than classical techniques (bayesian-based or geometric-based methods) in classical hyperspectral datasets \cite{chen2023improved}, as shown in applications (Section \ref{sec:application}). They are therefore not considered in this work.

Thus, for the choice of the classifier, as the captured data is generally not labelled, we focus here on unsupervised approaches only. More specifically, we consider the data as being distributed into distinguishable clusters in space: we therefore use classical clustering algorithms, such as the $k$-means algorithm or Gaussian mixture models (GMM), for which an associated space partition gives the classification.

Once the classifier chosen, an appropriate spatial density function must be defined. Regular approaches are developed in the second part, where we show that they suffer from important limits. In this paper, we propose a different and simple spatial density function addressing these limits, based on the Euclidean distance to convex polyhedra defined by any linear classifier, which allows building an appropriate abundance or probability map, adapted to any type of spectral images, whether Eq.\ref{eq:YMA} is suitable or not.

We show in this paper that our approach, in addition to being generalized to any kind of spectral images (even grayscale ones) regarding any chosen linear classifier, can surpass state-of-the-art methods - geometrical and even deep learning ones typically built to solve Eq.\ref{eq:YMA} - in terms of density map reconstruction, by applying it on the famous Samson hyperspectral dataset, which is based on the endmember-mixture modelling. Its application to an original multispectral dataset of Lithium-ion battery clearly not suitable to Eq.\ref{eq:YMA} demonstrates its generality.


\section{Density Functions}
\label{sec:methods}

\subsection{Problem with Regular Approaches}

\noindent Usual spatial density functions for clustering are distance-based functions. Typically, for the $k$-means algorithm with $K$ classes, the spatial probability $P$ associated with the cluster $k$ is a function of the distance $d$ to the computed cluster's centroid $c_k$: it can be the softmax function of the opposite distances 

\begin{figure}[hbt]
\centering
    \begin{subfigure}{.36\textwidth}
      \includegraphics[width=.95\linewidth]{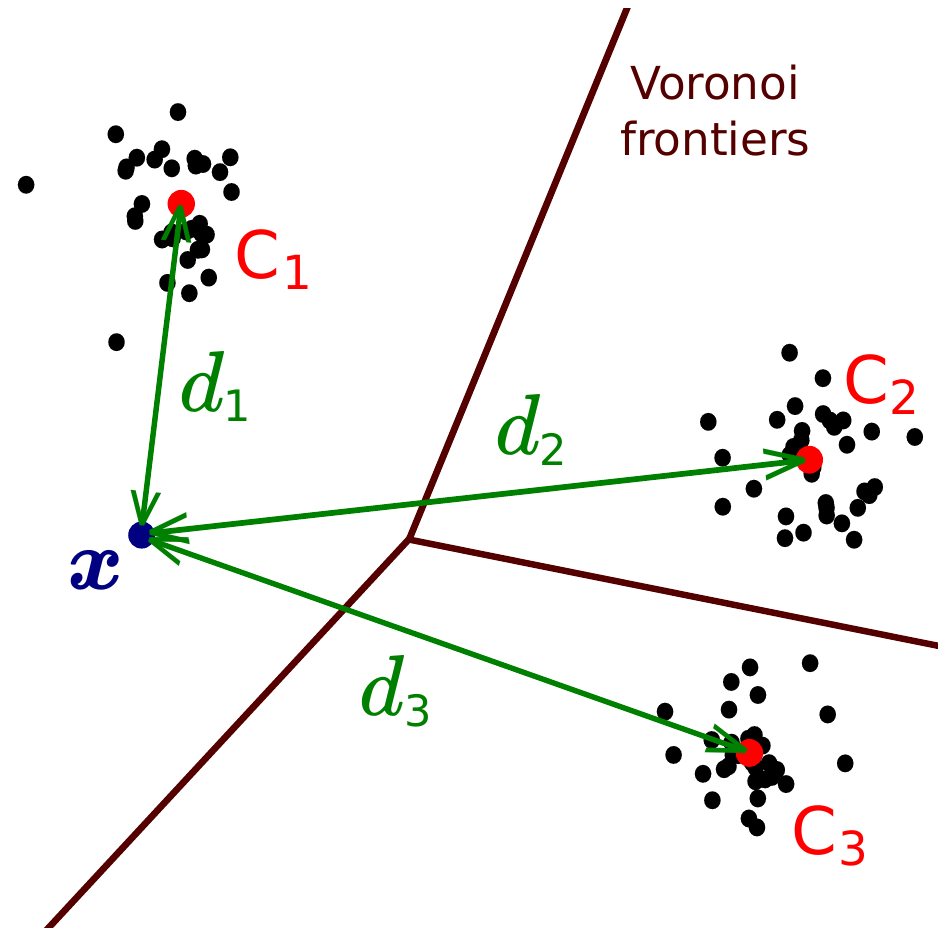}
      \caption{Distances to centroids.}
      \label{fig:centers1}
    \end{subfigure}%
    \hspace{15mm}
    \begin{subfigure}{.36\textwidth}
      \includegraphics[width=.95\linewidth]{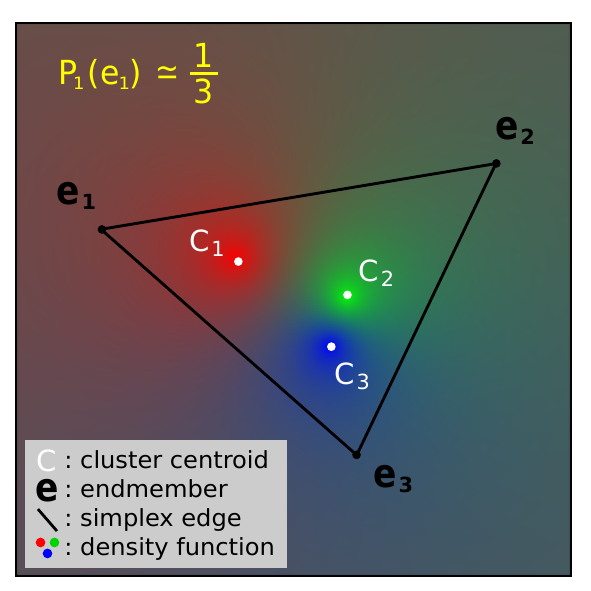}
      \caption{Density functions.}
      \label{fig:centers2}
    \end{subfigure}
    \caption{The distances to clusters' centers (red in \ref{fig:centers1}, white in \ref{fig:centers2}) given by the $k$-means algorithm (\ref{fig:centers1}) are used to compute density functions (color map in \ref{fig:centers2}) using Eq.\ref{eq:inverse_distance}: endmembers have lower density values than the center of their corresponding class (\ref{fig:centers2}).}
    \label{fig:centers}
\end{figure}

\vspace{2.0mm}
\begin{equation}
    P_k(x) = \frac{\text{exp}(- \alpha \hspace{.003\textwidth} d(x,c_k))}{\sum_{i=1}^{K}{\text{exp}(- \alpha \hspace{.003\textwidth} d(x,c_i))}}
    \label{eq:softmax}
\end{equation}
\vspace{2.0mm}

\noindent with $\alpha > 0$ the smoothing parameter (usually, $\alpha = 1$), or the normalised inverse distance function (Fig.\ref{fig:centers}) 

\vspace{2.0mm}
\begin{equation}
    P_k(x) = \frac{d(x,c_k)^{-p}}{\sum_{i=1}^{K}{d(x,c_i)^{-p}}}
    \label{eq:inverse_distance}
\end{equation}
\vspace{2.0mm}

\noindent (if $x \neq c_i$ $\forall i \in \llbracket 1, K \rrbracket$) with $p > 0$ the power parameter (usually, $p = 1$) \cite{chang2006parameter}.

For GMM, we can either use the Gaussian mixture function, which is already a density function, or use the probability functions above (\ref{eq:softmax}, \ref{eq:inverse_distance}) applied to the Mahalanobis distances of the Gaussian distributions. 

Although these density functions are widely used and appropriate in a context of classification only, one major issue is that they are not compatible with the endmember-mixture modelling (Eq.\ref{eq:YMA}):

\begin{itemize}
    
    \item the presented distance-based density functions give a higher value to points close to the centers (means) of the clusters than to any further point ; thus, as the data is inside a simplex for which the vertices represent the endmembers, these last ones will have a lower density value than the centers computed by any clustering algorithm (Fig.\ref{fig:centers},\ref{fig:centers2}) ;
    
    \item the Gaussian mixture function and the Mahalanobis distance-based approach (the GMM modelling alone) do not guarantee path-connected class subsets ; therefore, two different endmembers of the simplex could be associated with the same class (or at least have close density vectors).
    
\end{itemize}

\begin{figure}[t]
\centering
    \begin{subfigure}{.25\textwidth}
      \includegraphics[width=.925\linewidth]{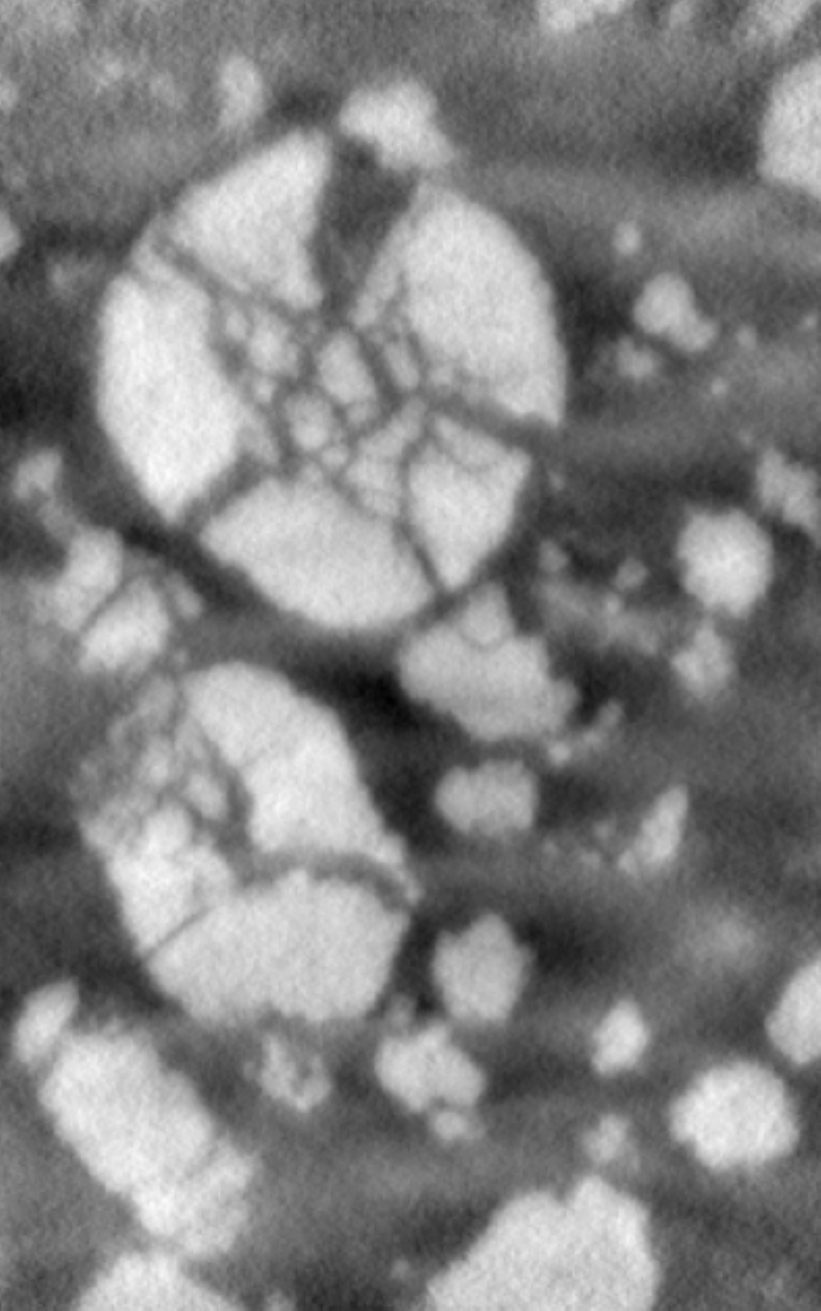}
      \caption{Original image.}
      \label{fig:holes1}
    \end{subfigure}%
    \hspace{10mm}
    \begin{subfigure}{.25\textwidth}
      \includegraphics[width=.925\linewidth]{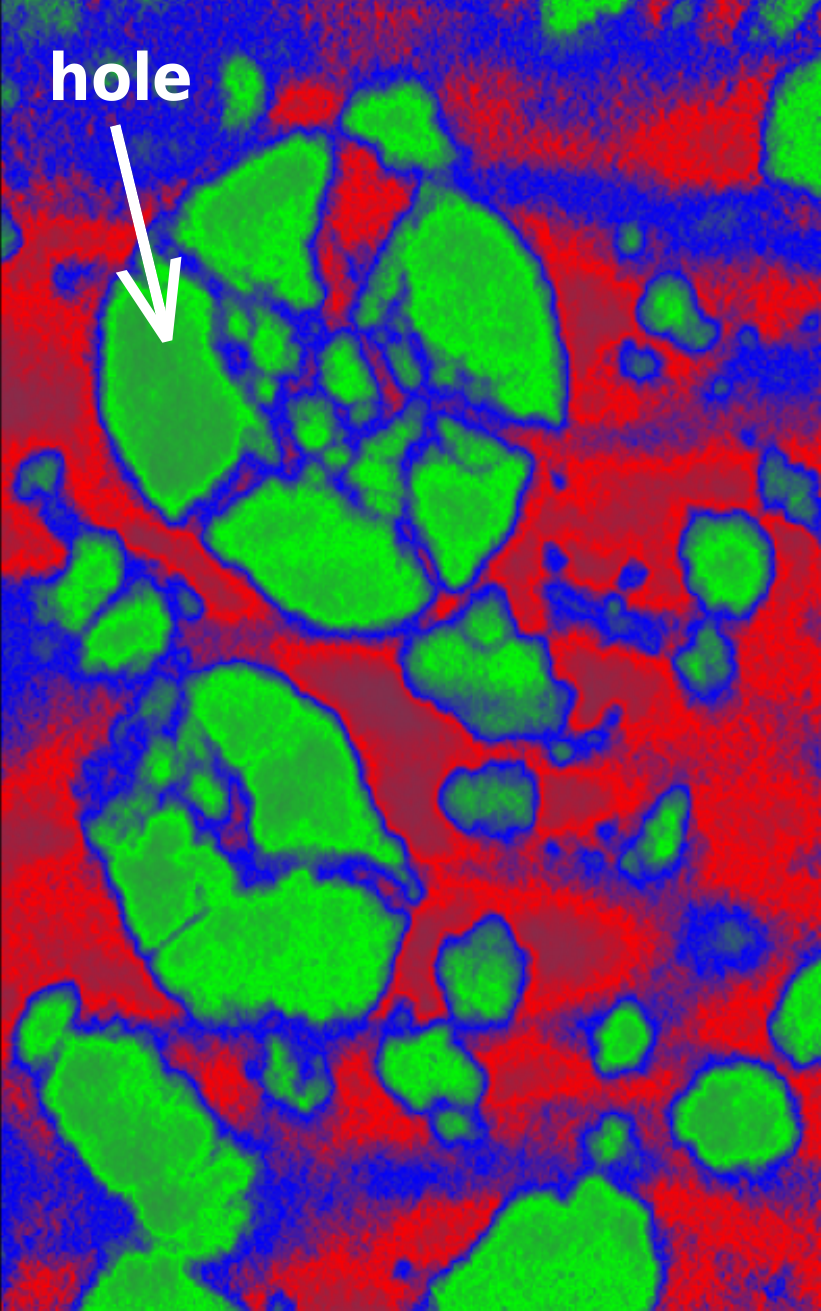}
      \caption{Computed map.}
      \label{fig:holes2}
    \end{subfigure}%
    \hspace{10mm}
    \begin{subfigure}{.25\textwidth}
      \includegraphics[width=.925\linewidth]{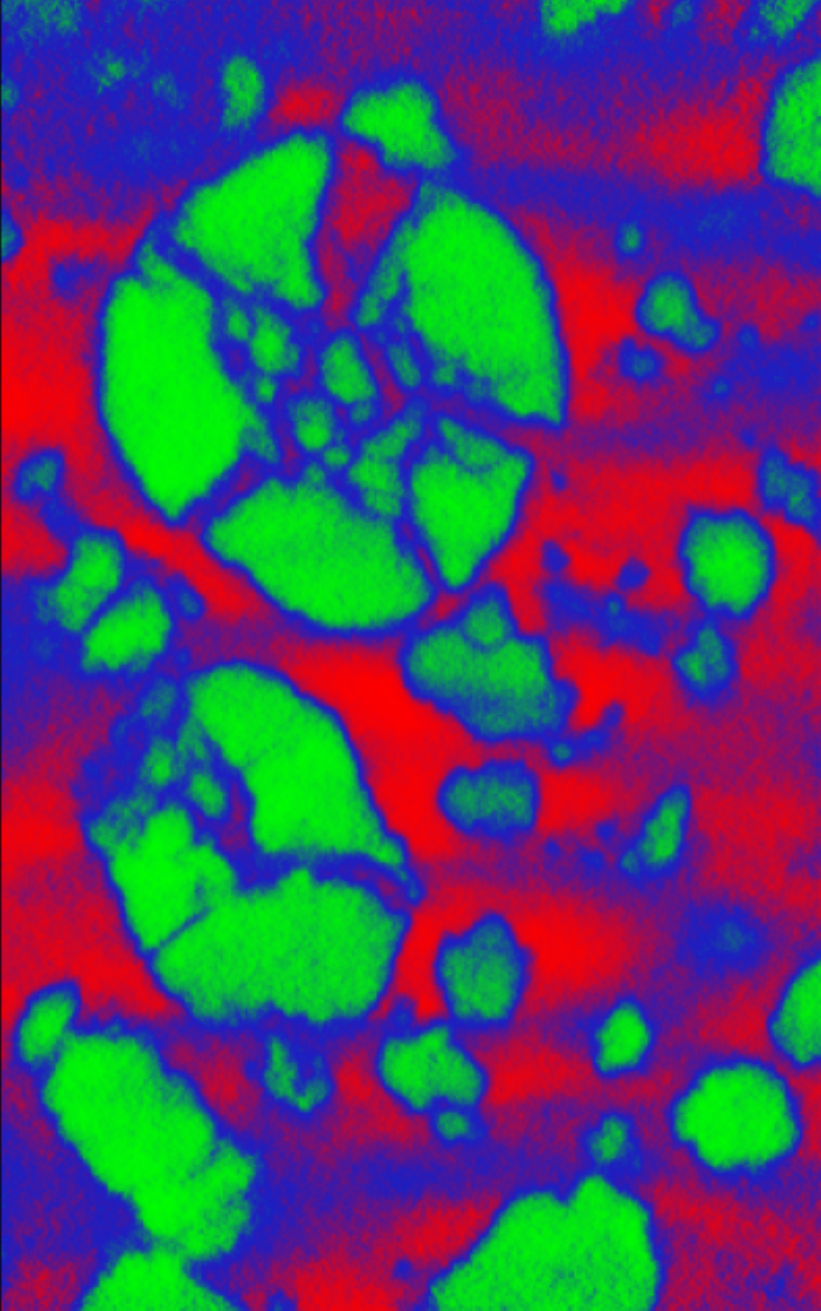}
      \caption{Expected map.}
      \label{fig:holes3}
    \end{subfigure}
    \caption{Grayscale image from the first band of a four-band spectral image of a Lithium-ion battery (\ref{fig:holes1}), and probability map computed with Eq.\ref{eq:inverse_distance} on three $k$-means centroids (\ref{fig:holes2}) to segment three chemical phases (high, medium and low values in \ref{fig:holes1}): the density function creates holes in the probability map compared to the expected one (\ref{fig:holes3}) ; see green phase.}
    \label{fig:holes}
\end{figure}

The same observations can be made for the other usual density functions associated with clustering algorithms, such as the Fuzzy C-means or any other functions based on the distance to clusters or to their centers. One consequence of such density functions is that \say{holes} appear in density maps: in Fig.\ref{fig:holes} hereinafter representing the electrochemical medium of a Lithium-ion battery with three visible chemical phases (red, green, blue), as the centers of clusters given by such clustering algorithms are not located on the extreme spectral values, the usual density functions assign a lower probability value to the brightest pixels than to pixels closer to the center of the class, regarding the brightest class (cracking particles in green). 

\vspace{4.0mm}
\vspace{2.0mm}

\vspace{2.0mm}
\vspace{2.0mm}

We thus need to define a density function which guarantees path-connected classes, such that the more a point is \say{deep} inside its class (or \say{far} from the others), the highest density value it will be assigned to. This way, with an adapted classifier (or space partitioning), the endmembers (and points close to them) would have the highest density values, and contrasts inside the classes will be preserved.


\subsection{Proposed Approach}

The idea of the proposed approach is quite simple: instead of taking distances to clusters (or their centers), we will consider signed distances to the classes' frontiers. Moreover, if the data is suitable, to guarantee that the classes are path-connected, and to make the structure more harmonious and the interpretation easier, we will consider here linear classifiers only.

\begin{figure}[t]
\centering
    \begin{subfigure}{.36\textwidth}
      \includegraphics[width=.95\linewidth]{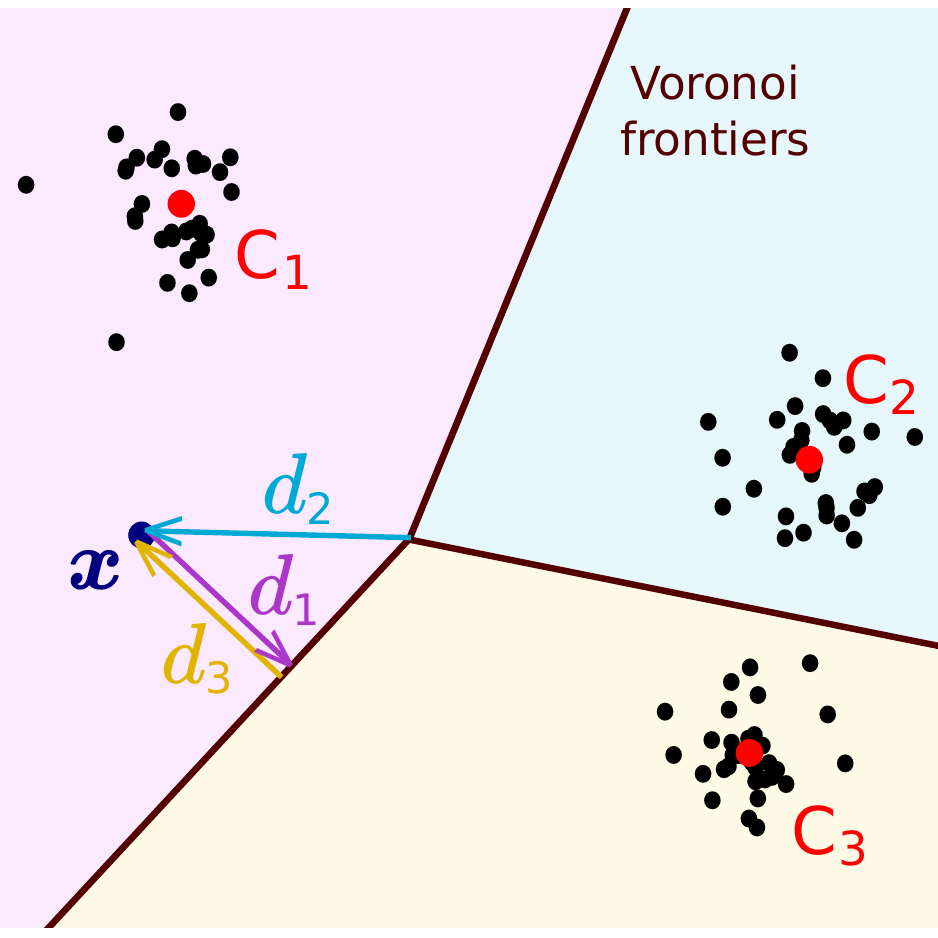}
      \caption{Signed distances to frontiers.}
      \label{fig:frontiers1}
    \end{subfigure}%
    \hspace{15mm}
    \begin{subfigure}{.36\textwidth}
      \includegraphics[width=.95\linewidth]{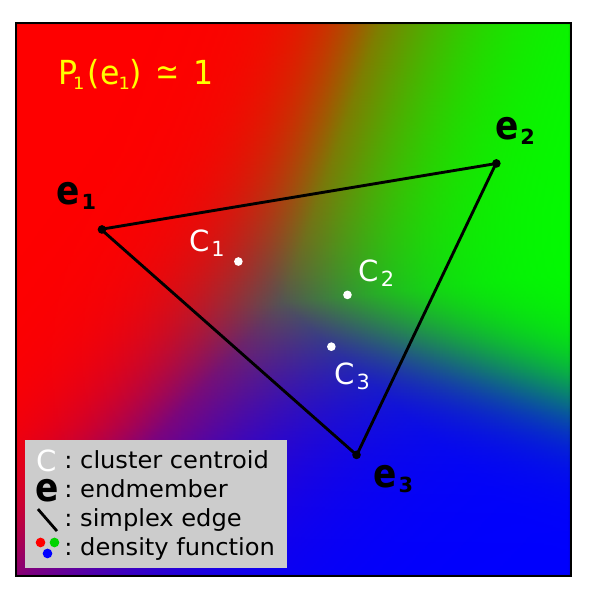}
      \caption{Density functions.}
      \label{fig:frontiers2}
    \end{subfigure}
    \caption{The signed distances to clusters' frontiers given by the $k$-means algorithm (\ref{fig:frontiers1}) are used to compute density functions (\ref{fig:frontiers2}) using Eq.\ref{eq:softmax}: endmembers have higher density values than the center (red in \ref{fig:centers1}, white in \ref{fig:centers2}) of their corresponding class (\ref{fig:frontiers2}).}
    \label{fig:frontiers}
\end{figure}

By definition of linear classifiers, frontier hyperplanes are built to separate the classes, which are then represented by distinct and complementary convex $n$-dimensional polyhedra in the Euclidean space $\mathbb{R}^n$. We thus consider the two following classifiers:

\begin{itemize}
    
    \item the $k$-means algorithm, for which the polyhedral classes are given by the corresponding Voronoi cells (Fig.\ref{fig:frontiers}) ;
    
    \item the application of a GMM on the unlabelled data, followed by the application of one-versus-one SVM to each pair of classes on the labelling given by the GMM, to build the frontier hyperplanes between polyhedral classes.
    
\end{itemize}

The first method is better adapted to isotropic Gaussian distributions with the same covariance matrix, the second one for anisotropic Gaussian distributions with different covariance matrices (more general). Other classifiers can obviously be used as long as they give polyhedral classes as results. 

The signed Euclidean distance between the measured point $x \in \mathbb{R}^n$ and each of the $k$ polyhedral classes $P \subseteq \mathbb{R}^n$ is then computed, resulting in a vector of $k$ signed distances associated with $x$, before applying the softmax function of the opposite distances (Eq.\ref{eq:softmax}) to obtain its corresponding density vector. 

\vspace{2.0mm}
\vspace{2.0mm}

Figure \ref{fig:frontiers} shows how this simple approach addresses the limits of existing density functions highlighted in the previous section: points which are the deepest ones in their respective class have the highest density values. If the true endmembers are not given, we can assume that, regarding the linear segmentation made, they are the most likely to be located in space where probabilities are the highest. Unlike regular approaches, it allows furthermore preserving the contrasts of segmented phases in the original spectral image inside their respective classes (Fig.\ref{fig:holes3}).

Although the distance from a point in the Euclidean space to a convex polyhedron is easy to represent, computing it turns out to be a challenging task. In the following sections, we formulate the problem mathematically, review the different approaches in the literature, and highlight the problems raised by the existing methods. We propose in this paper a mathematical process allowing computing the exact minimum-norm point to any convex polyhedron, which, despite of having an exponential complexity in the worst case along the number $k$ of support hyperplanes of the polyhedron (iif $n > 2$), turns out to be fast in practice for one to thirty hyperplanes.


\section{Distance to Convex Polyhedra: Problem Formulation}
\label{sec:problem}

\subsection{Definitions and Mathematical Formulation}

\noindent We work here in the Euclidean space $\mathbb{R}^n$ of finite dimension $n \in \mathbb{N}^*$ with inner product $\langle \cdot , \cdot \rangle$. To each of the $k$ (affine) hyperplanes $H \subseteq \mathbb{R}^n$ given by any linear classifier, is associated one unique scalar-vector couple $(s,v) \in \mathbb{R} \times \mathbb{R}^n$, where $v$ is the normed vector orthogonal to $H$ and $s$ the distance of $H$ to the zero point $0_n$ signed by direction $v$, such that 

\vspace{1.0mm}
\begin{equation}
    H = \{ x \in \mathbb{R}^n \hspace{0.5mm}|\hspace{0.5mm} \langle x , v \rangle = s \}.
    \label{eq:hyperplane}
\end{equation}
\vspace{-0.5mm}

\noindent As $v$ is unique, $H$ is oriented. Such an hyperplane $H$ is the frontier of a unique closed (affine) halfspace $B$ \say{behind} $H$ regarding $v$ such that 

\vspace{-0.5mm}
\begin{equation}
    B = \{ x \in \mathbb{R}^n \hspace{0.5mm}|\hspace{0.5mm} \langle x , v \rangle \leq s \}.
    \label{eq:halfspace}
\end{equation}
\vspace{-0.5mm}

\noindent We write $H_{(s,v)}$ (resp. $B_{(s,v)}$) the hyperplane defined by Eq.\ref{eq:hyperplane} (resp. halfspace defined by Eq.\ref{eq:halfspace}) regarding $(s,v)$. Polyhedra can then be properly defined. 

\vspace{1.5mm}
\begin{definition}[Polyhedron]
A subset $P \in \mathbb{R}^n$ is a (closed convex) \textit{polyhedron} if it is the intersection of a finite number of closed halfspaces. It can be either bounded or unbounded. A bounded polyhedron is called polytope. 
\normalfont{\cite{bruns2009polytopes}}
\label{def:polyhedron}
\end{definition}
\vspace{1.5mm}

Let $h \in \mathcal{F}(I, \mathbb{R} \times \mathbb{R}^n)$ be a family of $k$ scalar-vector couples $h_i = (s_i,v_i) \in \mathbb{R} \times \mathbb{R}^n$ indexed by $I = \llbracket 1 , k \rrbracket$. We write $P_h$ the polyhedron defined as follows 

\vspace{-0.75mm}
\begin{equation}
    P_h = \bigcap_{i \in I} B_{h_i} .
    \label{eq:polyhedron}
\end{equation}
\vspace{0.5mm}

\begin{figure}[ht]
\centering
    \begin{subfigure}{.36\textwidth}
      \includegraphics[width=.95\linewidth]{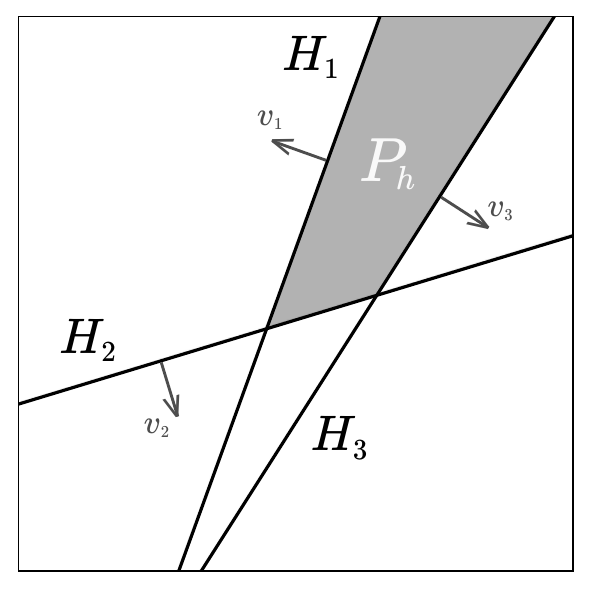}
      \caption{A polyhedron.}
      \label{fig:polyhedron}
    \end{subfigure}%
    \hspace{15mm}
    \begin{subfigure}{.36\textwidth}
      \includegraphics[width=.95\linewidth]{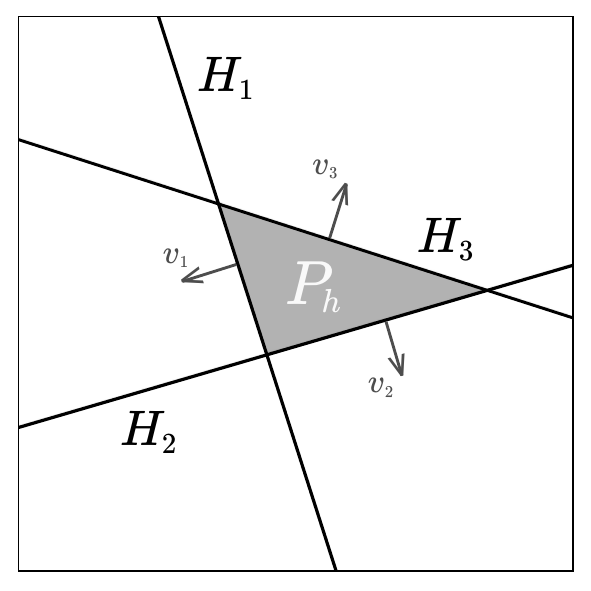}
      \caption{A polytope (bounded).}
      \label{fig:polytope}
    \end{subfigure}
    \caption{Example of an unbounded polyhedron (\ref{fig:polyhedron}) and of a bounded one (\ref{fig:polytope}) formed by three closed halfspaces.}
    \label{fig:poly_example}
\end{figure}

\noindent Figure \ref{fig:poly_example} hereinafter allows us to better visualize this. 

\vspace{2.0mm}

This representation of a convex polyhedron as the intersection of closed halfspaces (Eq.\ref{eq:polyhedron}) is called a $H$-representation \cite{grunbaum1967convex}. Alternatively, but for polytopes only, polyhedra can be represented as the convex hull of a finite set of points in $\mathbb{R}^n$, which are its vertices. This less general representation is called the $V$-representation of a polyhedron. 

The distance function $d$ between two points $x,y \in \mathbb{R}^n$ is the usual Euclidean distance: $d(x,y) = {\|x-y\|}_2$. The distance between a point $x \in \mathbb{R}^n$ and a subset $S$ of $\mathbb{R}^n$ is then defined as 

\vspace{1.5mm}
\begin{equation}
    \begin{array}{crcl}
    d\colon & \mathbb{R}^n \times \mathcal{P}(\mathbb{R}^n) & \rightarrow & \mathbb{R_+} \\
     & x \hspace{0.5mm},\hspace{0.5mm} S & \mapsto & \text{inf}_{y \in S}{{\|x-y\|}_2} 
    \end{array} .
    \label{eq:distance}
\end{equation}
\vspace{-0.5mm}

\vspace{1.5mm}
\begin{definition}[Minimum-norm point]
A point $y$ in the subset $S \subseteq \mathbb{R}^n$ minimizing the Euclidean distance to $x \in \mathbb{R}^n$ is called a minimum-norm point in $S$ from $x$. 
\normalfont{\cite{fujishige1990KJ00001202729}}
\label{def:min_norm_point}
\end{definition}
\vspace{1.5mm}

\noindent If $S$ is convex, then the minimum-norm point $y \in S$ from the fixed point $x \in \mathbb{R}^n$ is unique. 

The challenge here is, given $h \in \mathcal{F}(I, \mathbb{R} \times \mathbb{R}^n)$, to find a way of determining the minimum-norm point in the convex polyhedron $P_h \subseteq \mathbb{R}^n$ from any $x \in \mathbb{R}^n$. This problem is also known in the literature as the nearest point problem in a polyhedral set \cite{liu2012nearest}. If $x \in P_h$ - which can be easily verified by checking if $\max_{i \in I}{\{\langle x , v_i \rangle - s_i\}}$ is non-positive -, then the nearest point in $P_h$ is $x$ itself. 

As we are looking for the signed distance from $x$ to the polyhedron's frontiers $\partial P_h$ in the objective of computing our proposed density function using Eq.\ref{eq:softmax}, we define the usual signed distance function $d_s$ between a point $x \in \mathbb{R}^n$ and a subset $S$ of $\mathbb{R}^n$ as 

\vspace{-1.0mm}
\begin{equation}
    \begin{array}{crcl}
    d_s\colon & \mathbb{R}^n \times \mathcal{P}(\mathbb{R}^n) & \rightarrow & \mathbb{R} \\
     & x \hspace{0.5mm},\hspace{0.5mm} S & \mapsto & \text{sgn}(\mathds{1}_{x \notin S} - \frac{1}{2}) \times d(x, \partial S) 
    \end{array} .
    \label{eq:signed_distance}
\end{equation}
\vspace{0.75mm}

If $x \in P_h$, then its signed distance to the frontiers of $P_h$ is its distance to the complementary $P_h^\complement$ of $P_h$ in $\mathbb{R}^n$ put to the negative ; otherwise, it is the distance to the polyhedron itself. In this first case, $d_s$ (Eq.\ref{eq:signed_distance}) can be easily put under an explicit formula: $\forall x \in P_h ,$ 

\vspace{-1.75mm}
\begin{equation}
    d_s(x,P_h) = \max_{i \in I} \frac{\langle x , v_i \rangle - s_i}{{\|v_i\|}_2} .
    \label{eq:dPh_neg}
\end{equation}
\vspace{0.75mm}

\noindent In the second case ($x \notin P_h$), where the signed distance $d_s$ to $P_h$ is the Euclidean distance $d$ to $P_h$, there is unfortunately no explicit usable formula for the general case. Bergthaller and Singer managed to give an exact expression of the solution, but which uses undetermined parameters \cite{BERGTHALLER1992111}. To compute it, we need an algorithmic approach. In the next section, we review the different existing approaches from the state-of-the-art, and show their limits from a mathematical point of view.


\subsection{Existing Algorithms and their Limits}

In the literature, most of the methods designed to solve the nearest point problem in a polyhedral set are geometric-based approaches, where a polyhedron $P$ is seen as a geometrical structure in space, defined either by its vertices ($V$-representation) or by its support hyperplanes ($H$-representation), and where the problem is solved using projection-based algorithms. 

Equivalently, in $H$-representation, we can consider this problem as a convex quadratic programming problem, where $P$ is the set of all solutions $p \in \mathbb{R}^n$ to the linear matrix inequality $V p \leq S$, with $V$ the matrix of all the $v_i$ and $s$ the vector of all the $s_i$ in the space centered on the reference $x \in \mathbb{R}^n$, and where the minimum-norm point $y \in \mathbb{R}^n$ is given as a solution to

\vspace{1.0mm}
\begin{equation}
\begin{array}{ll}
    \vspace{0.2mm}
    \text{Minimize} & {\|y\|}_2^2 \\
    \text{Subject to} & Vy \leq s 
\end{array} .
\label{optim}
\end{equation}
\vspace{1.0mm}

Among the classical methods, Wolfe's algorithm \cite{wolfe1976finding} and Fujishige's dual algorithm \cite{fujishige1990KJ00001202729} remain popular processes for finding the minimum-norm point in a convex polytope. Several other algorithms have been developed for the three-dimensional case only \cite{dyllong1999accurate}, which is too restrictive for our problem. 

Most of the methods that use $H$-representation only consider the convex quadratic programming problem (\ref{optim}) solved using conventional algorithms, such as the simplex method \cite{simplexmethod}, interior point methods \cite{goldfarb1990n}, successive projection methods \cite{ruggiero2000modified}, or the Frank-Wolfe algorithm \cite{frank1956algorithm}.

Recent algorithms either revisit these classical methods \cite{jaggi2013revisiting}, are based on complex objects that require a relatively large amount of computational effort \cite{liu2012nearest}, or are based on a gradient descent such as the Operator Splitting Quadratic Program (OSQP) \cite{osqp}. 
Regardless of the approach, the known methods can be classified into two main categories: 

\begin{itemize}

    \item the ones which guarantee to give the exact solution to the problem in a finite number of iterations (Wolfe, Fujishige, Dyllong, etc.) ;
    
    \item the ones which give, unless in particular cases, an approximation of the solution only, based on an iterative algorithm converging to the optimal solution for an infinite number of iterations (Frank-Wolfe, Interior Point, OSQP, etc.).
    
\end{itemize}

In our case, even though it is not a necessary requirement, we will focus on exact methods only. Moreover, as the distance must be computed for all the pixels in our spectral images and to each of the polyhedral classes, we need to run the algorithm thousands to millions of times depending on the size of the image: we therefore need a fast and light algorithm for our practical case, i.e. where polyhedra are defined by a small or medium number of hyperplanes. 

Most of the exact methods are based on the vertices, which is critical for us, as linear classifiers usually return the family $h$ of the couples $(s,v)$ describing separation hyperplanes between polyhedral classes ($H$-representation), and as there necessarily are unbounded polyhedra in the resulting segmented space. Converting a polyhedron into its $V$-representation is computationally expensive, as we have first to verify that it is unbounded ($k \geq n$), then to find all its vertices, resulting in $\binom{k}{n}$ equations to solve \cite{saaty1955number}. 

We therefore developed a geometric-based algorithm which uses some mathematical properties of polyhedral sets to optimize the iterative research of the minimum-norm point, and which is fast in practice for a small to medium number of hyperplanes defining the polyhedron (one to thirty). In the next section, we'll present the properties on which relies the algorithm and its main lines.


\section{An Exact Minimum-Norm Point Calculation Process for Convex Polyhedra}
\label{sec:process}

\subsection{Support Hyperplanes and Minimum \texorpdfstring{$H$}{H}-Description}

Before getting started with the algorithm and its properties, we will first simplify the problem. If the set of the halfspaces defining a polyhedron $P$ in its $H$-representation has not been processed yet - which is the case for the linear classifiers used in this work -, there may exist halfspaces which have no impact on the construction of $P$, i.e. for which their removal from the intersection (Eq.\ref{eq:polyhedron}) does not change set $P$. The objective of this subsection is to find a way of detecting all these \say{unnecessary} halfspaces to remove them from the intersection, to make $P$ lighter and have better performances on the proposed algorithm.

\vspace{1.5mm}
\begin{definition}[Support Hyperplane]
Let $H \subseteq \mathbb{R}^n$ be an affine hyperplane. If the polyhedron $P$ is contained in one of the two closed halfspaces bounded by $H$, then $H$ is called support hyperplane of $P$ if $P \cap H \neq \emptyset$.
\label{def:supportHP}
\end{definition}
\vspace{1.5mm}

\begin{figure}[ht]
\centering
    \begin{subfigure}{.36\textwidth}
      \includegraphics[width=.95\linewidth]{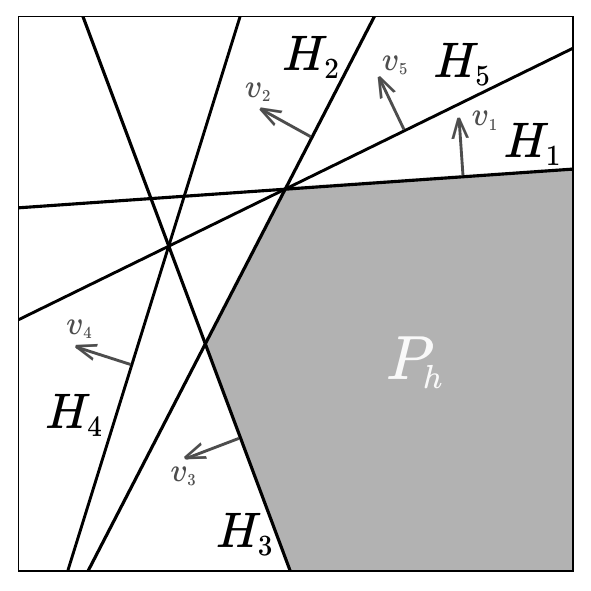}
      \caption{A polyhedron.}
      \label{fig:min_h1}
    \end{subfigure}%
    \hspace{15mm}
    \begin{subfigure}{.36\textwidth}
      \includegraphics[width=.95\linewidth]{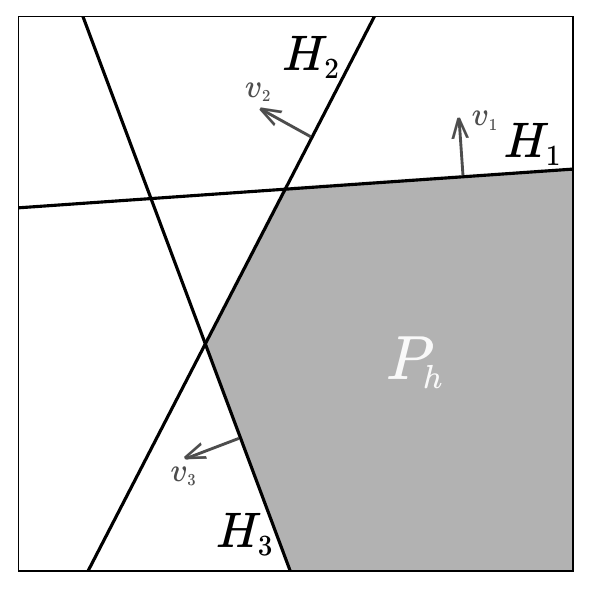}
      \caption{Its min $H$-description.}
      \label{fig:min_h2}
    \end{subfigure}
    \caption{Example of a polyhedron (\ref{fig:min_h1}) as the intersection of five halfspaces, and its minimum $H$-description (\ref{fig:min_h2}) where only the three first halfspaces have been preserved.}
    \label{fig:min_h}
\end{figure}

With such definition (\ref{def:supportHP}) given in \cite{bruns2009polytopes}, we can easily understand that all the halfspaces whose frontier hyperplane is not a support hyperplane of $P$ are unnecessary for the definition of $P$. In Figure \ref{fig:min_h} hereinafter, $H_4$ is not a support hyperplane of $P$: the corresponding halfspace is therefore removed from the intersection defining $P$. 

Verifying $P \cap H \neq \emptyset$ for all hyperplanes $H$ of the description of a polyhedron $P$ is very simple and allows removing part of the unnecessary halfspaces. But the condition of being support hyperplanes is actually not sufficient to remove all the unnecessary halfspaces to obtain what we call the \say{minimum $H$-description} (\ref{def:minHD}) of $P$ \cite{grunbaum1967convex}: in Fig.\ref{fig:min_h1}, $H_5$ is a support hyperplane as it \say{touches} one of the polyhedron's vertices, but is not necessary, because, if we remove its associated halfspace from the intersection, $P$ does not change. We thus need a more powerful filtering condition on halfspaces.

\vspace{1.5mm}
\begin{definition}[Minimum $H$-description]
Let $\mathcal{B} = (B_1, B_2, \ldots, B_k)$ be a family of $k$ closed halfspaces, and let $P = \bigcap_{i \in I} B_i$. We call minimum $H$-description of $P$ a subfamily $\mathcal{B}'$ of $\mathcal{B}$ with $k'$ elements, such that $\bigcap_{i \in I'} B_i' = P$ and $\forall j \in I'$, $ \bigcap_{i \in I' \setminus \{j\}} B_i' \neq P$.
\label{def:minHD}
\end{definition}
\vspace{1.5mm}

\noindent Note that if $P$ is full-dimensional (i.e. of dimension $n$), then its minimum $H$-description is unique. 

From this definition (\ref{def:minHD}) comes the following proposition, which allows directly verifying if an halfspace $B_j$ in the description of $P$ is in its minimum $H$-description (thus is necessary) or not.

\vspace{1.5mm}
\begin{proposition}
The minimum $H$-description of $P$ is the family of halfspaces $B_j$ in $\mathcal{B}$ such that $B_j^\complement \cap \left( \bigcap_{i \in I \setminus \{j\}} B_i^{\mathrm{o}} \right) \neq \emptyset$.
\label{prop:minHD}
\end{proposition}
\vspace{1.0mm}

\noindent $B^\complement$ being the complement of $B$ in $\mathbb{R}^n$, and $B^{\mathrm{o}}$ its interior.

As the condition in Proposition \ref{prop:minHD} uses open sets only (complement and interior of a closed set), it can be easily expressed by a condition on a strict linear matrix inequality, with $j \in I$, as follows

\vspace{0.0mm}
\begin{equation}
    \exists x \in \mathbb{R}^n, \hspace{2.0mm} A^j x < b^j
    \label{eq:minHD}
\end{equation}
\vspace{-1.5mm}

\noindent with the matrix $A^j = (v_1, \ldots, v_{j-1}, -v_j, v_{j+1}, \ldots, v_k)^\intercal$ and the vector $b^j = (s_1, \ldots, s_{j-1}, -s_j, s_{j+1}, \ldots, s_k)^\intercal$.

The condition in Proposition \ref{prop:minHD} is equivalent to verifying the consistency of the matrix inequality in (\ref{eq:minHD}). To do so, regular approaches can be used, such as linear programming with the zero function to minimize subject to the constraints (\ref{eq:minHD}), or the $I$-rank of the system for small dimensions of $A^j$ \cite{dines1919systems}. This way, each time we'll need to, we can easily get the minimum $H$-description of any polyhedron $P$.

\subsection{Preliminary Algorithm}

The main algorithm which allows computing the exact minimum-norm point in any polyhedral set $P$ is based on successive projections of the reference $x$ on methodically-chosen support hyperplanes of $P$ until the minimum-norm point is reached. 

To better understand its main steps and the properties it uses, we will first introduce a preliminary algorithm $A_1$ (\ref{alg:1}), which allows projecting the reference point $x \in \mathbb{R}^n$ on an intersection of hyperplanes. 

\begin{algorithm}[ht]
    \SetKwInOut{Input}{Input}
    \SetKwInOut{Output}{Output}

    \Input{$\cdot$ $x \in \mathbb{R}^n$\\ $\cdot$ $h \in \mathcal{F}(I, \mathbb{R} \times \mathbb{R}^n)$ such that $(v_i)_{i\in I}$ is linearly independent}
    \Output{ $y \in \mathbb{R}^n$ such that $y \in \cap_{i \in I} H_i$}
    
    $y \gets x$ \tcp*[r]{The moving point $y$ is initialized as the reference point $x$}
    $U \gets \emptyset$ \tcp*[r]{The set $U$ of orthonormal vectors is initialized as empty}
    \For{$i \in I$}{
        $w_i \gets v_i - \sum_{u\in U} \langle v_i, u \rangle u$ \tcp*[r]{We remove from $v_i$ its projections on the vectors $u$ from $U$}
        $u_i \gets \frac{w_i}{\|w_i\|}$ \tcp*[r]{We normalize $w_i$ (has non-zero norm by linear indep.of $(v_i)_i$)}
        $d_i \gets \frac{\langle x , v_i \rangle - s_i}{\langle u_i, v_i \rangle}$ \tcp*[r]{We compute the distance $d_i$ in direction $u_i$ between $y$ and $H_i$}
        $y \gets y - d_i u_i$ \tcp*[r]{We move $y$ in direction $u_i$ of distance $d_i$ such that $y$ is in $H_i$}
        $U \gets U \cup \{u_i\}$ \tcp*[r]{We add the new vector $u_i$ to the set $U$ of orthonormal vectors}
    }
    
    \caption{Projection on the intersection of hyperplanes.}
    \label{alg:1}
\end{algorithm}

From any $x \in \mathbb{R}^n$ and any $h \in \mathcal{F}(I, \mathbb{R} \times \mathbb{R}^n)$ for which the $v_i$ are linearly independent, algorithm $A_1$ successively projects $x$ on all the hyperplanes formed by couples $(s_i,v_i)$ in $h$ in the order given by $I$. At each iteration $i \in I$, with the aim of projecting $x$ on the $i$-th hyperplane, the projection direction $w_i$ is computed from $v_i$ by removing from it all its components $\langle v_i, u \rangle u$ in the space formed by the set of previous projection directions $U$. This way, by moving $x$ at iteration $i$ in direction $w_i$ (which is non-zero, as the $v_i$ are linearly independent), the point will always stay in the hyperplanes considered in previous iterations $j < i$, as $w_i$ is orthogonal to all the previous $v_j$ by construction. Then, $w_i$ is normed, resulting in $u_i$, and the projection distance $d_i$ is computed such that moving $x$ in direction $u_i$ with a distance $d_i$ will allow the new $x$ being in the $i$-th hyperplane. $u_i$ is then added to $U$ before going to the next iteration. Note that $U$ is formed by the Gram-Schmidt process on the family $(v_i)_{i \in I}$.

This simple algorithm has fundamental properties which will be used for the main algorithm. Let's now consider any family $h \in \mathcal{F}(I, \mathbb{R} \times \mathbb{R}^n)$ of $k$ scalar-vector couples $h_i = (s_i,v_i) \in \mathbb{R} \times \mathbb{R}^n$ indexed by $I$. We have the following proposition.

\vspace{1.5mm}
\begin{proposition}
Let $h^*$ be a subfamily of $h$ such that the $v_i^*$ are linearly independent. Then, the result of $A_1$ $y = A_1(x,h)$ is the minimum-norm point in the intersection $\cap_{i \in I^*} H_i^*$ of hyperplanes formed by $h^*$ from $x$.
\label{prop:minInter}
\end{proposition}
\vspace{1.5mm}

As it is obvious that, if $x \notin P_h$, there exist at least one support hyperplane of $P_h$ such that the minimum-norm point in $P_h$ from $x$ is in this hyperplane, from Proposition \ref{prop:minInter} directly comes the following corollary.

\vspace{1.5mm}
\begin{corollary}
There exists a subfamily $h^\dagger$ of $h$ such that $A_1(x,h^\dagger)$ is the minimum-norm point in $P_h$ from $x$.\\ 
\vspace{-2.0mm}

\noindent Writing $h'$ the subfamily of all the couples $h_i$ of $h$ whose hyperplane $H_i$ contains the min-norm point in $P_h$ from $x$, $h^\dagger$ is more precisely any subfamily of $h'$ such that $(v_i)_{i \in I^\dagger}$ is a basis of $\text{span}((v_i)_{i \in I'})$, with $I^\dagger$ the indices on $h^\dagger$ and $I'$ the ones on $h'$. 
\label{cor:minRela}
\end{corollary}
\vspace{1.5mm}

With Corollary \ref{cor:minRela}, we are facing the one problem: how may we determine such a subfamily $h^\dagger$? Before this, how may we determine $h'$, i.e. the family of the hyperplanes containing the minimum-norm point? 

As we know, such subfamilies are actually hard to determine without having any information on the polyhedron's vertices, or without using complex and computationally-expensive structures \cite{liu2012nearest}. Our method consists then in modifying algorithm $A_1$ to search the minimum-norm point in $P_h$ by recursively projecting $x$ on all the possible hyperplane combinations, until the min-norm point is reached.

\subsection{Main Algorithm}

As the number of hyperplane combinations is an exponential function of the number $k$ of support hyperplanes of $P_h$ ($2^k$), we then need a methodical search: we want to avoid unnecessary combinations, and start the search with the hyperplanes that are the most likely to contain the minimum-norm point. 

The first thing that we modify in $A_1$ is the addition of dimension reduction at each iteration: as, from iteration $i$ to $i+1$, the new direction vector $u_{i+1}$ and the new distance of projection $d_{i+1}$ are built such that $x$ stays in the hyperplanes previously considered, we will, each time we go \say{deeper} in the projections from $i$ to $i+1$, instead of considering the problem in $\mathbb{R}^n$, consider it in the affine subspace of lower dimension formed by the hyperplane $H_i$ on which has just been projected $x$, and transform the family $h$ into a new one $h'$ which is expressed in this new subspace as follows

\vspace{1.5mm}
\begin{equation}
    \begin{array}{cc}
    \forall j \in I \setminus \{i\}, \hspace{-1.5mm} & 
    \left\{
    \hspace{-0.5mm}
    \begin{array}{rcl}
        \vspace{1.0mm}
        v_j' \hspace{-1mm}&\hspace{-1mm} = \hspace{-1mm}&\hspace{-1mm} \frac{v_j - \langle v_j, v_i \rangle v_i / {\|v_i\|}_2^2}{{\left\| v_j - \langle v_j, v_i \rangle v_i / {\|v_i\|}_2^2 \right\|}_2} \\
        s_j' \hspace{-1mm}&\hspace{-1mm} = \hspace{-1mm}&\hspace{-1mm} \langle x , v_j' \rangle - \frac{\langle x , v_j \rangle - s_j}{\langle v_j, v_j' \rangle} 
    \end{array}
    \right. .
    \end{array}
    \label{eq:reduction}
\end{equation}
\vspace{1.5mm}

This way, at each new iteration $i+1$, after reducing space dimension and the family $h$ into $h'$ using Eq.\ref{eq:reduction}, the new direction vector $u_{i+1}'$ will simply be $v_i' / {\|v_i'\|}_2$, and the distance of projection $d_{i+1}'$ will be the signed distance from $x$ to the halfspace $B_{h_{i+1}'}$, exactly as it is at iteration $1$ when $U$ is empty. 

This space reduction does not only allow working in a reduced space $\mathbb{R}^{n-i}$ at iteration $i+1$, but it also allows generalizing properties that can be made at iteration $1$ on $h$ to all the following iterations $i+1$ on the modified $h'$. Which is crucial for the following properties that will be used for the main algorithm. 

\vspace{1.5mm}
\begin{proposition}
There exists a couple $(s,v)$ in $h$ such that the signed distance $d_s$ between $x$ and the halfspace $B_{(s,v)}$ is positive, and its frontier hyperplane $H_{(s,v)}$ contains the minimum-norm point in $P_h$ from $x$.
\label{prop:positive}
\end{proposition}

\vspace{-2.0mm}
\begin{proposition}
If there exists a couple $(s,v)$ in $h$ such that the projection $x - \frac{d_s(x, B_{(s,v)})}{{\|v\|}_2} v$ of $x$ on the hyperplane $H_{(s,v)}$ is in $P_h$, then the signed distance $d_s$ between $x$ and $B_{(s,v)}$ is the maximum of the signed distances $d_s$ from $x$ to all the halfspaces defined by the couples in $h$, i.e.: $d_s(x, B_{(s,v)}) = \max_{i \in I} d_s(x, B_{(s_i,v_i)})$.
\label{prop:directProj}
\end{proposition}

\vspace{0.0mm}
\begin{proposition}
If $n \leq 2$ and $h$ describes the min $H$-description of $P_h$, then the min-norm point in $P_h$ from $x$ is in the hyperplane of maximum (positive) distance to $x$, i.e. in $H_{(s_i,v_i)}$ where $i = \text{argmax}_{i \in I} {d(x,B_{(s_i,v_i)})}$.
\label{prop:2D}
\end{proposition}

\vspace{1.0mm}
\noindent Note that the reciprocals of Prop. \ref{prop:directProj} and \ref{prop:2D} are false.

\vspace{1.5mm}
\begin{criterion}
$y \in P_h$ is the min-norm point in $P_h$ from $x$ if and only if 
\hspace{0.5mm}$P_h^{\mathrm{o}} \cap B_{(s^*,v^*)}^{\mathrm{o}} = \emptyset$\hspace{0.5mm}, 
with $s^* = \langle y, y-x \rangle$ and $v^* = y-x$.
\label{prop:criterion}
\end{criterion}
\vspace{1.5mm}

Criterion \ref{prop:criterion} allows verifying if a point $y$ in $P_h$ is the min-norm point from $x$, and is equivalent to the consistency of a strict linear matrix inequality like (\ref{eq:minHD}).

The main algorithm is a recursive algorithm which can either go \say{deeper} in the projection process of $x$ considering allowed-to-project hyperplanes, or go back in a previous state of $x$ if deeper projections are not possible or unnecessary at the current recursion. At the beginning of each recursion, we consider the whole original family $h$, which is then filtered, transformed and sorted using the propositions seen before. At the end of the recursion, the algorithm enters a loop over the filtered $h_i$ in which $x$ is projected on $H_i$ and then put in a deeper recursion step. This way, $x$ can be projected on all the possible combinations of hyperplanes, but which are methodically filtered and sorted, until the min-norm point is reached. 

\vspace{2.5mm}
\noindent Each recursion then follows these main lines:

\begin{enumerate}

    \item if $x$ is the minimum-norm point in $P_h$ (Criterion \ref{prop:criterion}), then stop and return $x$ ; otherwise, continue ;

    \item if $x \in P_h$ but is not the minimum-norm point, then go back in the previous recursion ;

    \item transform $h$ into the reduced $h'$ using Eq.\ref{eq:reduction} ;
    
    \item filter $h'$:
    \begin{itemize}
        \item remove from $h'$ the $h_i'$ whose $v_i'$ is linearly dependant of $U$ (set of orthonormal projection vectors from the previous recursions) for Corollary \ref{cor:minRela} ;
        \item remove from $h'$ the $h_i'$ for which $d_s(x,B_{h_i}) \leq 0$ (Prop. \ref{prop:positive}) ;
        \item compute the minimum $H$-description of $P_{h'}$ (useless $h_i'$ are removed from $h'$) using Eq.\ref{eq:minHD};
    \end{itemize}
    
    \item if the filtered $h'$ is empty, then go back in the previous recursion ;

    \item sort $h'$ from the greatest distance $d_s(x,B_{h_i'})$ to the smallest, to increase the chances of projecting first on a hyperplane containing the minimum-norm point (Prop. \ref{prop:directProj} and \ref{prop:2D}) ;
    
    \item in a loop, for $h_i'$ in $h'$:
    \begin{itemize}
        \item project $x$ on $H_{h_i'}$ ;
        \item definitely remove $h_i'$ from $h'$ for the next deepest recursions in the current loop (it avoids permutations) ;
        \item call the function with these new parameters ;
        \item if the minimum-norm point has not been found, put back $x$ in its previous state ;
    \end{itemize}
    
    \item if the minimum-norm point has not been found yet, go back in the previous recursion ;
    
\end{enumerate}

\subsection{Comparison with a State-of-the-Art Algorithm}

Built this way, our algorithm ensures to return the exact solution to the nearest point problem in a polyhedron $P$, in a finite number of steps. However, even if the search of the minimum-norm point is optimized by mathematical properties developed in the previous subsections, its complexity in the worst case is exponential, in $\mathcal{O}(2^k)$ times a polynomial expression of $k$, with $k$ the number of support hyperplanes of $P$.

To evaluate its complexity in time over the number $k$ in practical case, we implemented it in the C programming language, and built a stochastic model which generates polyhedra with a given number $k$ of support hyperplanes in dimension $n$. We then chose to compare the performances of our algorithm to one of the most recent methods able to rapidly solve convex quadratic problems such as ours: the OSQP solver \cite{osqp}. This solver is based on an automatically-optimized gradient descent, and computes an approximation of the solution point.

\begin{figure}[ht]
\centering
    \includegraphics[width=.6\linewidth]{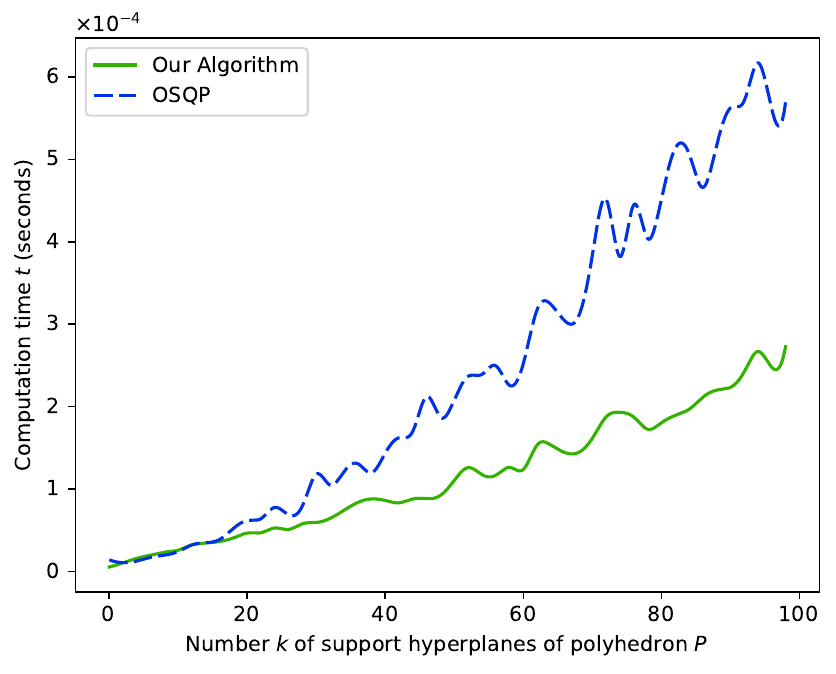}
    \caption{Evolution of computation time $t$ over the number $k$ of support hyperplanes (averaged over $1000$ simulations per value of $k$), for the OSQP solver (blue) and our algorithm (green), in dimension $n = 3$.}
    \label{fig:graphN3}
\end{figure}
\vspace{2.0mm}

\begin{figure}[!ht]
\centering
    \includegraphics[width=.6\linewidth]{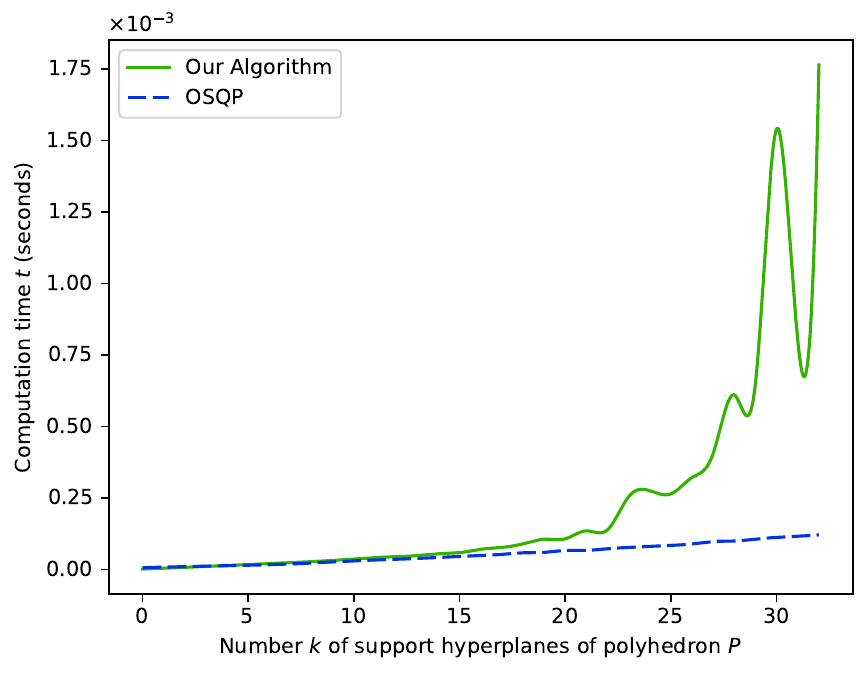}
    \caption{Evolution of computation time $t$ over the number $k$ of support hyperplanes (averaged over $1000$ simulations per value of $k$), for the OSQP solver (blue) and our algorithm (green), in dimension $n = k$.}
    \label{fig:graphNM}
\end{figure}
\vspace{0.5mm}

The two figures Fig.\ref{fig:graphN3} and Fig.\ref{fig:graphNM} are the results of two experiences: in the first one (\ref{fig:graphN3}), we fixed $n = 3$ to study the performances of the algorithms in the $3$-dimensional Euclidean space ; in the second one (\ref{fig:graphNM}), we let $n = k$ (as, in spectral images, the number of bands, $n$, is usually greater than the number of classes or endmembers, $\geq k+1$). Both graphs represent the evolution of computation time over the number $k$. The blue discontinuous curve representing the OSQP solver, with a given relative tolerance of $10^{-6}$, and the green continuous one representing our method. For every $k$, the computation time values are the means of $1000$ simulations with different polyhedra.

Figure \ref{fig:graphN3} reveals that, when $n$ is fixed and relatively small ($n=3$), our method has similar or even better performances than the OSQP solver, and this for all values of $k$. It moreover seems that, the greater $k$ is, the more significant the difference between OSQP and our algorithm is. The expected exponential behaviour of our method actually appears here to be almost linear or polynomial, from $k=1$ to $k=100$.

On the other hand, Figure \ref{fig:graphNM} highlights a way different behaviour of our algorithm when $n$ follows $k$: its associated computation time stays equivalent to the OSQP's one from $k=1$ to around $k=15$ or $20$, but becomes exponential over $k$ after $k=20$ and explodes around $k=30$. This shows that our algorithm is limited when we need to consider more than $30$ support hyperplanes in higher dimensions. In practice, as we use it for our density function in cases where the number of considered classes rarely exceeds twenty or thirty, this behaviour will not be a problem for us.

\vspace{1.5mm}
\begin{table}[ht]
\caption{Mean and standard deviation of the distance between the point computed by the OSQP solver and the exact point from our method (over all the $K \times 1000$ simulations).}\label{tab:distances} \centering
\begin{tabular}{|c|c|c|}
  \hline
  \textbf{Experience} & $n = 3$ & $n = k$ \\
  \hline
  \hline
  \textbf{Mean error} & $9.87 \times 10^{-3}$ & $6.31 \times 10^{-4}$ \\
  \hline
  \textbf{STD} & $1.08 \times 10^{-2}$ & $3.80 \times 10^{-4}$ \\
  \hline
\end{tabular}
\end{table}
\vspace{1.5mm}

As our method gives the exact solution and OSQP an approximation, one last thing to analyse is the mean distance between the point given by OSQP and the one by our method. Table \ref{tab:distances} shows that the mean error made by OSQP and the standard deviation are both higher in the case where $n=3$ than where $n=k$. In both cases, most of the distances are between $10^{-4}$ and $10^{-1}$ (sometimes greater when $n=3$), which may be not convenient if we look for a high precision.


\section{Application to Class Representation}
\label{sec:application}

\subsection{Abundance Map Estimation}

In this section, to evaluate the performances of our approach, we apply it to one of the most widely used hyperspectral datasets for spectral unmixing: the Samson dataset \cite{figliuzzi2016bayesian}. Typically well suited to the linear endmember-mixture modelling, it is composed of $156$ bands and represents $3$ regional classes: water, forest and soil. We choose here the use of a GMM followed by a SVM to segment space into polyhedral subsets. We then compute for each pixel $x$ its signed distance to each of the polyhedral classes using our exact algorithm previously introduced.

In this subsection, the idea is to determine a real abundance map $A$ (i.e. for which there exists some matrix $M$ such that $Y = M A$). To do so, we first determine the endmembers $M$ using our approach, to then retrieve the abundances $A$ using $A = M^{-1}Y$ (from Eq.\ref{eq:YMA}) like in classical approaches \cite{eches2011variational}. The challenge here is then the way of estimating $M$. To make it simple, and like most of the approaches, we assume that, among the observed data $Y$, there exist some spectra close enough to the ground-truth endmembers. To determine these endmembers $M$ using our method, we will take, inside each polyhedral class computed by our classifier over the observed data $Y$, the spectrum $Y_i$ which is the \say{deepest} one in the corresponding polyhedron, i.e. which has the lowest signed distance: it should then represent the \say{purest} class spectrum. 

\vspace{4.0mm}
\begin{figure}[hbt]
\centering
    \includegraphics[width=0.7\linewidth]{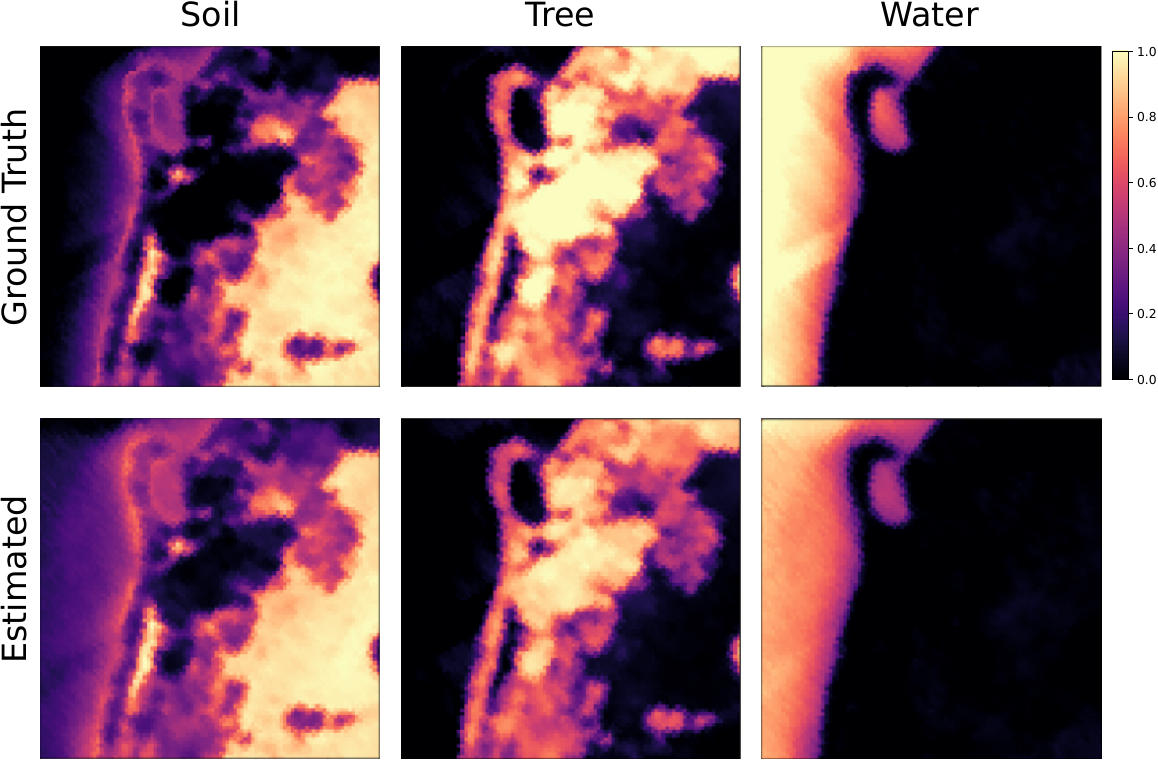}
    \caption{Ground truth ($1$st row) and estimated abundance maps ($2$nd row), for the three classes (columns) of Samson. The Root Mean Squared Error (RMSE) is about $0.1533$.}
    \label{fig:abundances}
\end{figure}

Figure \ref{fig:abundances} shows the abundance maps of the three classes given by our method (second row) compared to the ground truth (first row). To evaluate the quality of these results, we compare them to state-of-the-art methods among the most efficient ones for hyperspectral unmixing: a geometric distance-based approach like ours, the new maximum-distance analysis (NMDA) from \cite{tao2021endmember} ; and a deep learning approach, the spatial–spectral adaptive nonlinear unmixing network (SSANU-Net) from \cite{chen2023improved}. We computed the RMSE between our results and the ground truth, and compared it to the best RMSE of the abundances given in these papers \cite{tao2021endmember,chen2023improved} (Table \ref{tab:abundances}).

\vspace{1.5mm}
\begin{table}[ht]
\caption{RMSE of the abundance maps given by the three considered methods on the Samson dataset, and processing time (for ours, as GMM are stochastic models, RMSE and time are the means over $100$ runs).}\label{tab:abundances} \centering
\begin{tabular}{|c|c|c|c|c|}
  \hline
  \textbf{Method} & NMDA \cite{tao2021endmember} & SSANU-Net \cite{chen2023improved} & Ours (Abundance Map) & Ours (Probability Map) \\
  \hline
  \hline
  \textbf{RMSE [$A$]} & $0.1620$ & $0.1668$ & $0.1533$ & $0.0985$ \\
  \hline
  \textbf{time} (s) & $1.4743$ & unknown & $1.9697$ & $1.9528$ \\
  \hline
\end{tabular}
\end{table}
\vspace{1.5mm}

Table \ref{tab:abundances} shows that our approach (column \say{Ours (Abundance Map)}) has better performances than the two others (colums \say{NMDA \cite{tao2021endmember}} and \say{SSANU-Net \cite{chen2023improved}}) in terms of RMSE of the abundance map $A$. The last column (\say{Ours (Probability Map)}) corresponds to the results of our probability-map-like approach introduced in the next subsection. 

As our method depends on a probabilistic model (GMM), we averaged the RMSE and the computation time over $100$ independent runs, using a ratio of $0.2$ for the random sample extraction for the training of the GMM. These results given by our method seem quite stable, as the standard deviation of the RMSE on all the runs is about $0.0061$. 

The computation time (Tab.\ref{tab:abundances}) of our method is however greater than the NMDA's. But it is the addition of the training time of the GMM, the fitting time of the SVM, and the computation time of the distances to polyhedra given by our algorithm. Taken separately, the total computation time of our algorithm for all the pixels in the image and all the polyhedra is about $0.06$ s. only. 
Although these results are good, we have to remind that they highly depend on the classifier chosen.

\subsection{Probability Map Calculation}

We consider here the more general case, where we don't know whether the linear mixture modelling (Eq.\ref{eq:YMA}) is suitable to the spectral image or not. In this case, there is no search for endmembers or for linear combinations of them in the observed data: we simply use a given density function over the polyhedral partitioning of space made by a chosen classifier. 

We only use here the softmax function (Eq.\ref{eq:softmax}) of the opposite signed distances to polyhedral classes divided by their standard deviation. If the classes are not homogeneously shared in the spectral space - like in Samson's -, a change of basis can be made in the space of signed-distance vectors, using the vectors of lowest distance value for each class as new basis.

\vspace{4.0mm}
\begin{figure}[ht]
\centering
    \includegraphics[width=0.7\linewidth]{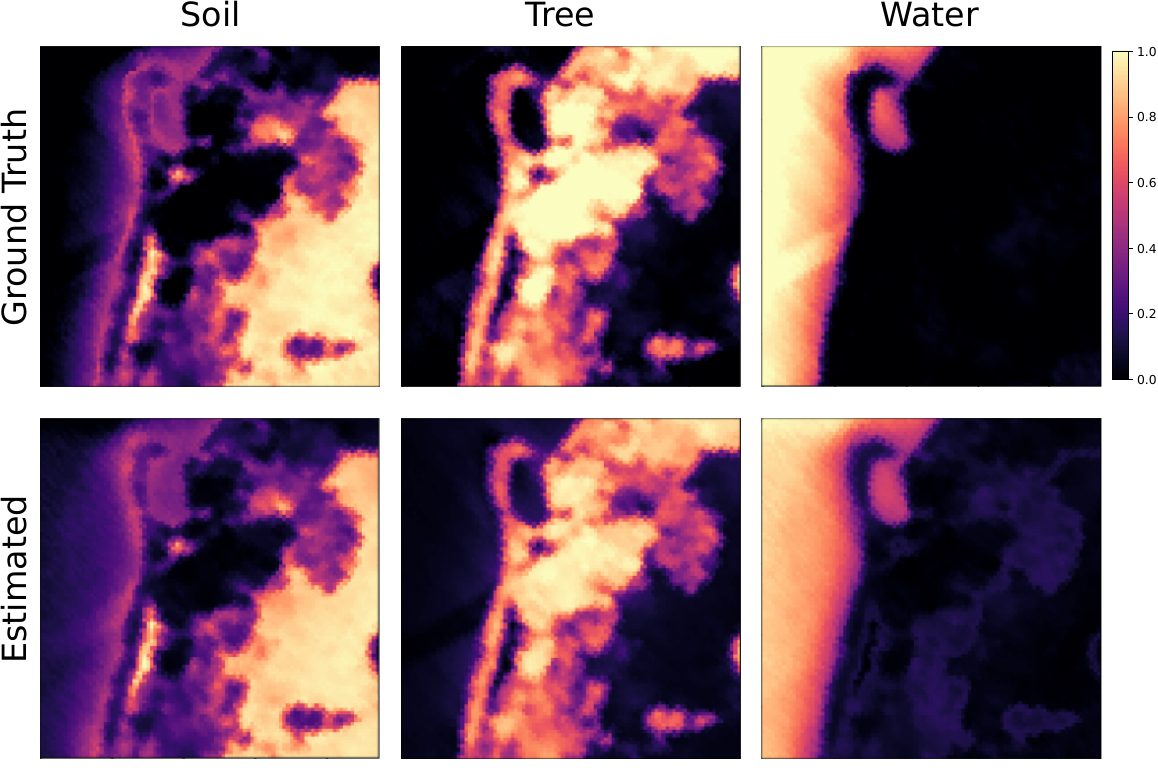}
    \caption{Ground truth ($1$st row) and computed probability maps ($2$nd row), for the three classes (columns) of Samson. The Root Mean Squared Error (RMSE) is about $0.0985$.}
    \label{fig:probabilities}
\end{figure}

The resulting density map $A$ is then, by construction, more suitable to a \textit{probability} map associated with the segmentation made (by here our GMM and SVM model) rather then an \textit{abundance} map, as it is not assured that there exists some matrix $M$ such that $Y = M A$ (Eq.\ref{eq:YMA}). We want to show, with the Samson dataset, that this general method still gives good results, even on spectral datasets which are well suited to Eq.\ref{eq:YMA}.

Figure \ref{fig:probabilities} reveals that this approach gives even a better estimation of the abundance maps on the Samson dataset than any of the linear unmixing approaches in table \ref{tab:abundances}: with the same parameters for the GMM, the mean RMSE (still over $100$ runs) between the determined maps and the ground truth is $0.0985$ (last column \say{Ours (Probability Map)}), with a standard deviation of $0.0103$, which represents the best results in terms of RMSE among the ones from our study and from the two papers taken as reference in this section, and probably one of the best results from the literature for the Samson dataset, regardless the approach.

\subsection{Phase Extraction}

In this last subsection, we want to validate our method by applying it on a dataset of spectral images which are clearly not suitable to the linear endmember-mixing modelling. 
To this end, we study here a set of spectral images of a Lithium-ion battery captured by X-ray nano-CT under four spectral bands (or \say{energies}). Our work originally started with this dataset. 

\vspace{4.0mm}
\begin{figure}[ht]
\centering
    \begin{subfigure}{.21\textwidth}
      \includegraphics[width=.95\linewidth]{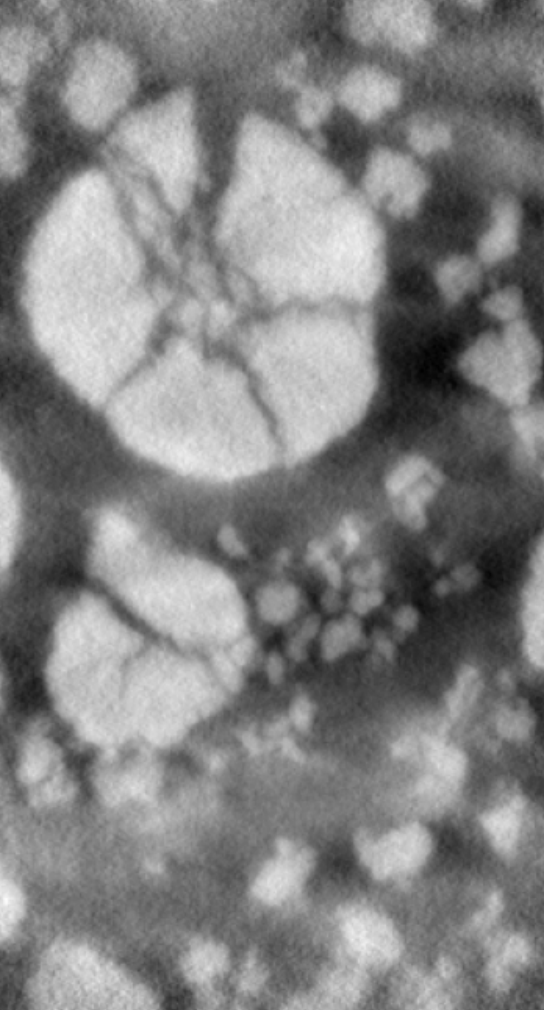}
      \caption{Band 1.}
      \label{fig:e1}
    \end{subfigure}%
    \hspace{5mm}
    \begin{subfigure}{.21\textwidth}
      \includegraphics[width=.95\linewidth]{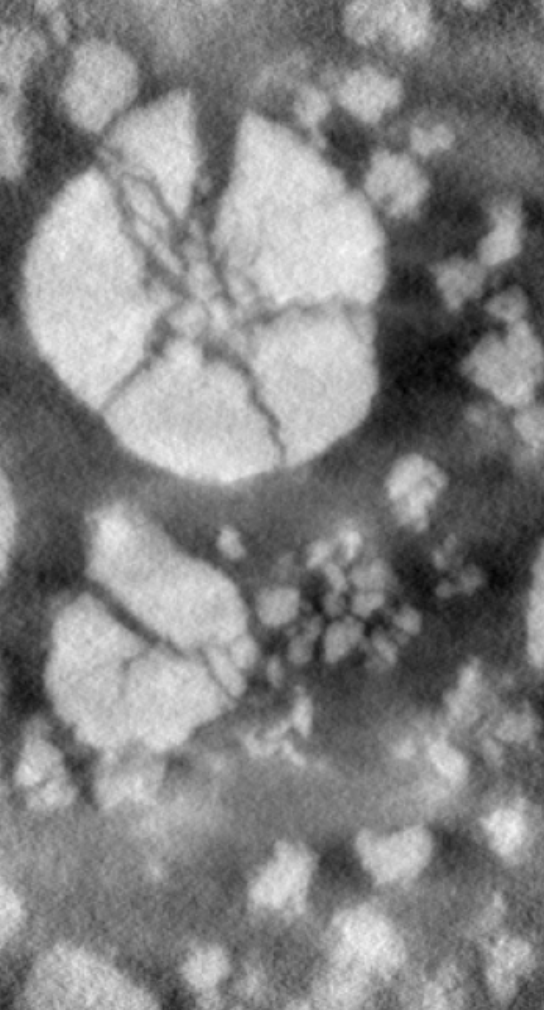}
      \caption{Band 2.}
      \label{fig:e2}
    \end{subfigure}%
    \hspace{5mm}
    \begin{subfigure}{.21\textwidth}
      \includegraphics[width=.95\linewidth]{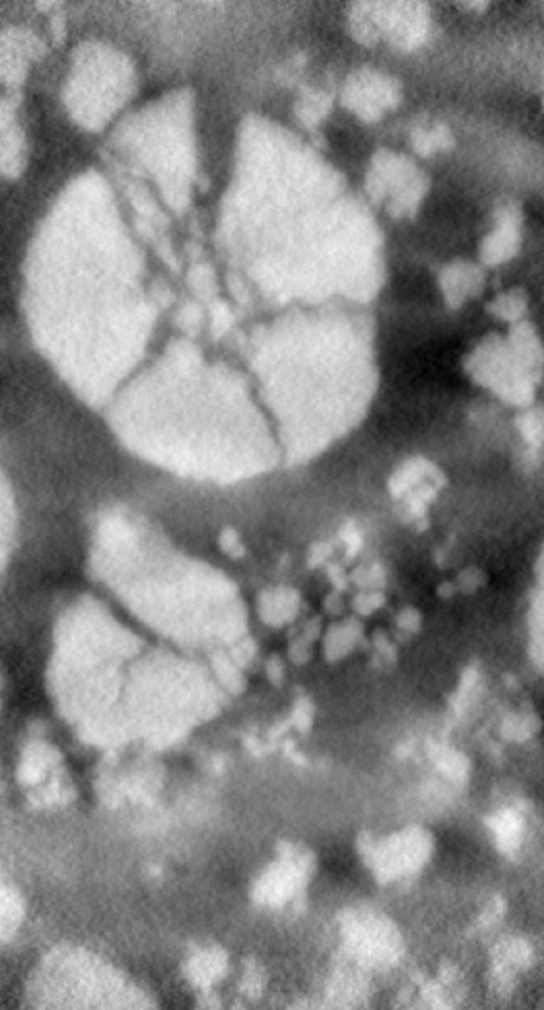}
      \caption{Band 3.}
      \label{fig:e3}
    \end{subfigure}%
    \hspace{5mm}
    \begin{subfigure}{.21\textwidth}
      \includegraphics[width=.95\linewidth]{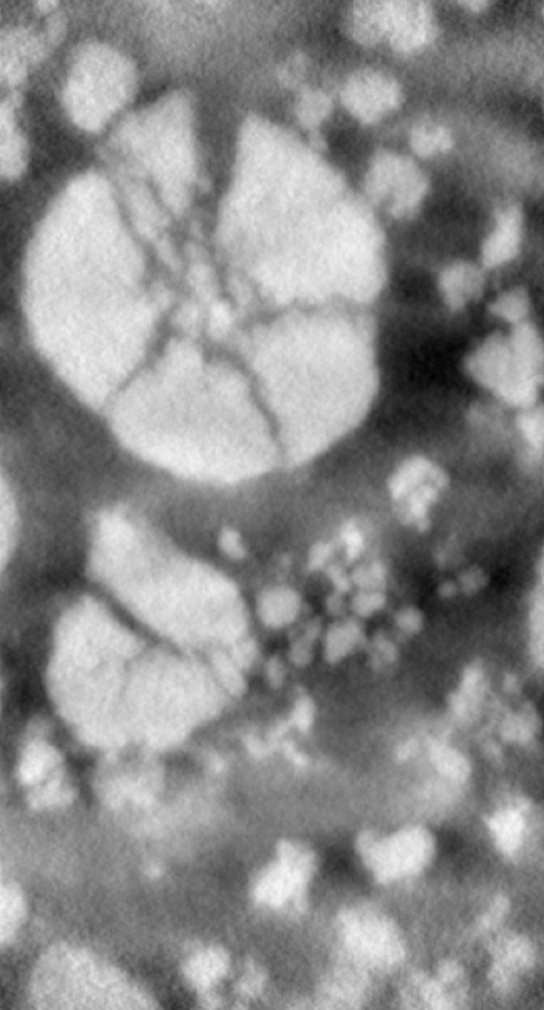}
      \caption{Band 4.}
      \label{fig:e4}
    \end{subfigure}
    \caption{Spectral image of a Lithium-ion battery captured by tomography X with four bands (\ref{fig:e1}, \ref{fig:e2}, \ref{fig:e3}, \ref{fig:e4}).}
    \label{fig:edges}
\end{figure}

\begin{figure}[!ht]
\centering
    \begin{subfigure}{.21\textwidth}
      \includegraphics[width=.925\linewidth]{grains_practice_e1.pdf}
      \caption{Original image (band 1).}
      \label{fig:a1}
    \end{subfigure}%
    \hspace{10mm}
    \begin{subfigure}{.21\textwidth}
      \includegraphics[width=.925\linewidth]{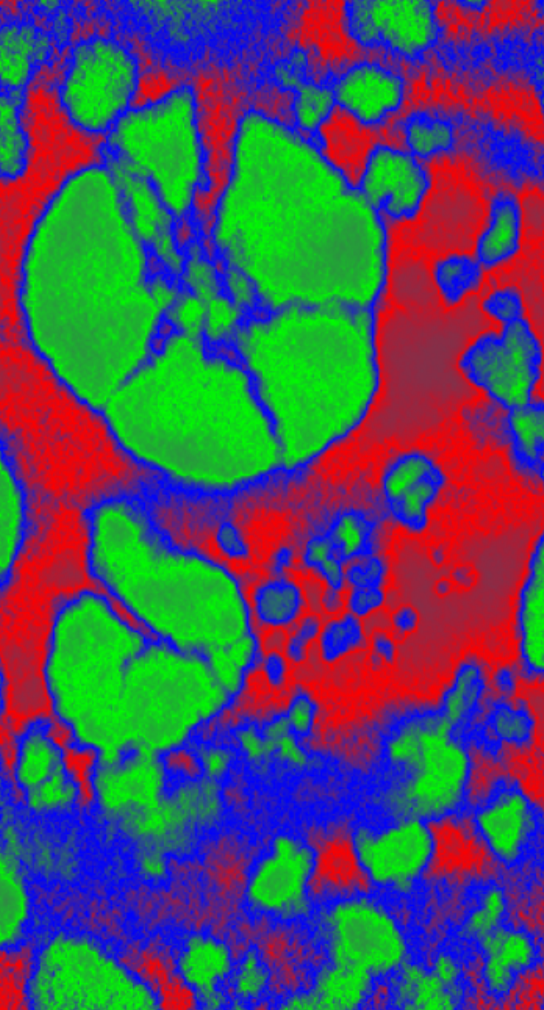}
      \caption{Usual probability map.}
      \label{fig:a2}
    \end{subfigure}%
    \hspace{10mm}
    \begin{subfigure}{.21\textwidth}
      \includegraphics[width=.925\linewidth]{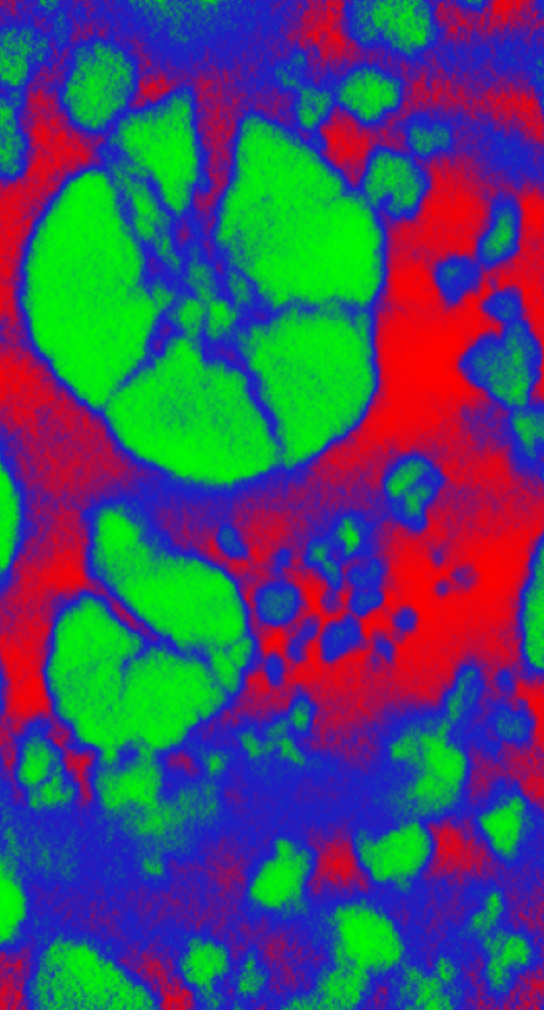}
      \caption{Our probability map.}
      \label{fig:a3}
    \end{subfigure}
    \caption{Image of the first band of the Li-ion battery dataset (\ref{fig:a1}), the usual probability map given by Eq.\ref{eq:inverse_distance} on the distances to $k$-means centroids (\ref{fig:a2}), and the one given by our approach with the same $k$-means parameters (\ref{fig:a3}): our method allows preserving contrasts in the phases, unlike the usual one in which holes are created (green phase).}
    \label{fig:results}
\end{figure}

Figure \ref{fig:edges} reveals how correlated the spectra of the data are: a pixel which has a certain value on one of the four bands is likely to have the same value on the other bands. Which is incompatible with any linear unmixing approach, as the data is distributed on one line in the spectral space, making the possible endmembers linearly dependant ($M$ not invertible). 

The objective is then to be able to extract the three visible phases in these images (Fig.\ref{fig:edges}): NMC particles (high values - green), CBD (blurry medium values - blue), and porosity (low values - red). These phases will be represented by their probability map determined by our general method previously seen, but with a $k$-means as classifier instead of a GMM. 

Unfortunately, there is no ground truth for this dataset to evaluate the results of our approach. But, in addition to mathematical guarantees, visual results in Fig.\ref{fig:results} allow validating the fact that our method preserves the contrasts (gradient) in probability maps inside the classes (\ref{fig:a3}). Which is not the case for usual density functions (\ref{fig:a2}). The visually-coherent map resulting from our approach validates its consistency in spectral images which cannot be linearly unmixed.


\section{Conclusions and Perspectives}
\label{sec:conclusion}

With the aim of addressing the cases where spectral images cannot be linearly unmixed, we developed a new approach which allows building an adapted density map from observed data. Density functions usually used for clustering models suffer from limits in the context of spectral unmixing: they are either based on the distances to clusters, which does not allow detecting any endmember and creates holes in density maps, or do not guarantee crucial spatial properties.

The new density function that we formulated to address these limits is based on the idea of computing the signed distance to the frontiers of polyhedral classes given by linear classifiers. We developed an algorithm capable of computing the exact minimum-norm point in any polyhedral subset. Despite its exponential complexity in the worst case, it remains faster than the recent OSQP solver \cite{osqp} in dimension $3$, and still finds the solution rapidly up to $30$ support hyperplanes in high dimension.

The application of our approach to the Samson dataset highlights a better estimation of the abundance maps than geometric-based and deep learning-based state-of-the-art approaches, whether in the context of abundance map or of probability map. In this last context, our method gives even much better results. Moreover, the results on a spectral dataset of a Li-ion battery, incompatible with linear unmixing approaches, validate its relevance in the general case. 

Despite such valuable results, some limits still remain: our algorithm for the minimum-norm point has an exponential behaviour in high dimension over $30$ hyperplanes, which is not desirable in practice for a great number of classes. Furthermore, testing the approach on other datasets, compatible with linear unmixing approaches or not, such as the Cuprite dataset \cite{tao2021endmember}, would bolster the observations and conclusions made on the studied datasets.

To go further, although we have focused solely on linear classifiers, we could extend our approach to non-linear methods by applying it in a space of higher dimension (feature map) given by a chosen mapping function, compute the minimum-norm points to polyhedral classes in it, before going back to the original space where classes and distances are non-linear.


\section*{Code Availability}

The full code is available at: \url{https://github.com/antoine-bottenmuller/polyhedral-unmixing}.

\vspace{2.0mm}
\noindent It contains:
\begin{itemize}
    \item the C and Python codes of the proposed algorithm for the minimum-norm point in a convex polyhedron ;
    \item the proposed density function based on the signed distance to polyhedral classes implemented in Python ;
    \item a complete example of the application of our two approaches (\textit{abundance} and \textit{probability}) to the Samson dataset (from: \url{https://rslab.ut.ac.ir/data}) in a Notebook file with commented results.
\end{itemize}


\section*{Acknowledgments}

This work was supported by the French Agence Nationale de la Recherche (ANR), project number ANR 22-CE42-0025.

\vspace{2.0mm}
\noindent We would also like to sincerely thank the three reviewers of the original article, published as a full paper in the Proceedings of the 14th ICPRAM (2025), for their thorough, insightful, and constructive comments. Their valuable feedback has helped us improve the clarity, depth, and overall quality of our manuscript.


\printbibliography 


\end{document}